\documentclass[twocolumn,spanish,english,amsmath,amssymb,aps,nofootinbib,superscriptaddress]{revtex4-1}
\usepackage[T1]{fontenc}
\usepackage[latin9]{inputenc}
\usepackage[a4paper]{geometry}
\usepackage{color}
\usepackage{babel}
\addto\shorthandsspanish{\spanishdeactivate{~<>}}
\usepackage{comment}
\usepackage{lipsum}
\usepackage{units}
\usepackage{halloweenmath}
\usepackage{mathrsfs}
\usepackage{amsmath}
\usepackage{amssymb}
\usepackage{graphicx}
\usepackage{esint}
\usepackage{xargs}[2008/03/08]
\usepackage[unicode=true,pdfusetitle,
 bookmarks=true,bookmarksnumbered=true,bookmarksopen=false,
 breaklinks=false,pdfborder={0 0 1},backref=false,colorlinks=false]
 {hyperref}
 \usepackage{caption}
\usepackage[most]{tcolorbox} 
\usepackage{lmodern} 
\usepackage[T1]{fontenc} 
\usepackage{float}
\usepackage{units}
\usepackage{graphicx}
\usepackage{esint}
\usepackage{mathrsfs}
\usepackage{dsfont}
\usepackage{caption}
\usepackage{placeins}

\hypersetup{ 
    colorlinks,
    linkcolor={},
    citecolor={},
    urlcolor={}
}

\begin{document}

\selectlanguage{spanish}%
\ifdefined\MacrosDosMilDieciocho
\else
\newcommand*{\MacrosDosMilDieciocho}{}

\newcommand{\DR}[1]{\textcolor{red}{[DR: #1]}}

\newcommand{\MS}[1]{\textcolor{blue}{[MS: #1]}}

\newcommand\blfootnote[1]{%
  \begingroup
  \renewcommand\thefootnote{}\footnote{#1}%
  \addtocounter{footnote}{-1}%
  \endgroup
}

\global\long\def\MODULO#1{\left|\,#1\,\right|}%

\global\long\def\PARENTESIS#1{\left(#1\right)}%

\global\long\def\CORCHETES#1{\left[#1\right]}%

\global\long\def\LLAVES#1{\left\{  #1\right\}  }%

\global\long\def\LKEY#1{\left\{  #1\right.}%

\global\long\def\RKEY#1{\left.#1\right\}  }%

\global\long\def\LCORCHETE#1{\left[#1\right.}%

\global\long\def\RCORCHETE#1{\left.#1\right]}%

\global\long\def\LPARENTESIS#1{\left(#1\right.}%

\global\long\def\RPARENTESIS#1{\left.#1\right)}%

\global\long\def\ANGULITOS#1{\left\langle #1\right\rangle }%

\global\long\def\ESPACIOLARGO{\hspace{10mm}}%

\global\long\def\ESPACIOMEDIO{\hspace{5mm}}%

\global\long\def\DEF{\overset{{\scriptscriptstyle \text{def}}}{=}}%

\global\long\def\LLAVEARRIBA#1#2{\overset{{\scriptstyle #2}}{\overbrace{#1}}}%

\global\long\def\LLAVEABAJO#1#2{\underset{{\scriptstyle #2}}{\underbrace{#1}}}%

\global\long\def\REALES{\mathbb{R}}%

\global\long\def\IMAGINARIOS{\mathbb{I}}%

\global\long\def\NATURALES{\mathbb{N}}%

\global\long\def\ENTEROS{\mathbb{Z}}%

\global\long\def\COMPLEJOS{\mathbb{C}}%

\global\long\def\RACIONALES{\mathbb{Q}}%

\global\long\def\DIFERENCIAL{\,\text{d}}%

\global\long\def\PRIME{{\vphantom{A}}^{\prime}}%

\global\long\def\HAPPY{\smiley}%

\global\long\def\SAD{\overset{\cdot.\cdot}{\frown}}%

\global\long\def\ORDER#1{\mathcal{O}\PARENTESIS{#1}}%

\global\long\def\DIRACDELTA#1{\delta\PARENTESIS{#1}}%

\newcommandx\KRONEDELTA[1][usedefault, addprefix=\global, 1=ij]{\delta_{#1}}%

\global\long\def\HEAVYSIDETHETA#1{\Theta_{H}\PARENTESIS{#1}}%

\newcommandx\LEVICIVITASYMBOL[1][usedefault, addprefix=\global, 1=ijk]{\varepsilon_{#1}}%

\global\long\def\SINC{\text{sinc }}%

\global\long\def\ATAN{\text{atan}}%

\global\long\def\INDICADORA#1{\mathbf{1}\left\{  #1\right\}  }%

\global\long\def\VECTOR#1{\boldsymbol{#1}}%

\global\long\def\VERSOR#1{\hat{\VECTOR{#1}}}%

\global\long\def\IDENTITY{\mathds{1}}%

\global\long\def\CURL{\VECTOR{\nabla}\times}%

\global\long\def\GRADIENT{\VECTOR{\nabla}}%

\global\long\def\DIVERGENCE{\VECTOR{\nabla}\cdot}%

\global\long\def\LAPLACIAN{\nabla^{2}}%

\global\long\def\REALPART#1{\text{Re}\left(#1\right)}%

\global\long\def\IMAGPART#1{\text{Im}\left(#1\right)}%

\global\long\def\TIENDEA#1{\underset{{\scriptscriptstyle #1}}{\longrightarrow}}%

\global\long\def\EVALUADOEN#1#2#3{\left\lceil #1\right\rfloor _{#2}^{#3}}%

\global\long\def\TERA#1{\text{ T}\unit{#1}}%

\global\long\def\GIGA#1{\text{ G}\unit{#1}}%

\global\long\def\MEGA#1{\text{ M}\unit{#1}}%

\global\long\def\KILO#1{\text{ k}\unit{#1}}%

\global\long\def\UNIT#1{\,\unit{#1}}%

\global\long\def\CENTI#1{\text{ c}\unit{#1}}%

\global\long\def\MILI#1{\text{ m}\unit{#1}}%

\global\long\def\MICRO#1{\text{ }\mu\unit{#1}}%

\global\long\def\NANO#1{\text{ n}\unit{#1}}%

\global\long\def\PICO#1{\text{ p}\unit{#1}}%

\global\long\def\FEMTO#1{\text{ f}\unit{#1}}%

\global\long\def\PORDIEZALA#1{\times10^{#1}}%

\global\long\def\PROBABILIDAD#1{\mathbb{P}\left(#1\right)}%

\global\long\def\COLOR#1#2{{\color{#2}{\,#1\,}}}%

\global\long\def\RED#1{\textcolor{red}{#1}}%

\global\long\def\BLUE#1{\COLOR{#1}{blue!80!white}}%

\global\long\def\GREEN#1{\textcolor{green!70!black}{#1}}%

\global\long\def\GRAY#1{\COLOR{#1}{black!30}}%

\global\long\def\GRAY#1{\COLOR{#1}{blue!35!white}}%

\global\long\def\GUNDERBRACE#1#2{\GRAY{\LLAVEABAJO{\COLOR{#1}{black}}{#2}}}%

\global\long\def\RUNDERBRACE#1#2{\RED{\LLAVEABAJO{\COLOR{#1}{black}}{#2}}}%

\global\long\def\BUNDERBRACE#1#2{\BLUE{\LLAVEABAJO{\COLOR{#1}{black}}{#2}}}%

\global\long\def\GUPBRACE#1#2{\GRAY{\LLAVEARRIBA{\COLOR{#1}{black}}{#2}}}%

\global\long\def\RUPBRACE#1#2{\RED{\LLAVEARRIBA{\COLOR{#1}{black}}{#2}}}%

\global\long\def\BUPBRACE#1#2{\BLUE{\LLAVEARRIBA{\COLOR{#1}{black}}{#2}}}%

\global\long\def\REDCANCEL#1{\RED{\cancel{{\normalcolor #1}}}}%

\global\long\def\BLUECANCEL#1{{\color{blue}\cancel{{\normalcolor #1}}}}%

\global\long\def\GREENCANCEL#1{\GREEN{\cancel{{\normalcolor #1}}}}%

\global\long\def\BLUECANCELTO#1#2{\BLUE{\cancelto{#2}{{\normalcolor #1}}}}%

\fi 

\global\long\def\UNITARYGROUP#1{\mathbf{U}\PARENTESIS{#1}}%

\global\long\def\DIRACCONJUGATE#1{\overline{#1}}%

\global\long\def\CONTRAVARIANTINDEX#1{\hspace{0cm}^{#1}}%

\global\long\def\COVARIANTINDEX#1{\hspace{0cm}_{#1}}%

\global\long\def\LAGRANGIAN{\mathscr{L}}%

\global\long\def\HAMILTONIAN{\mathscr{H}}%

\global\long\def\KET#1{\left|#1\right\rangle }%

\global\long\def\BRA#1{\left\langle #1\right|}%

\global\long\def\BRAKET#1#2{\ANGULITOS{\left.#1\right|#2}}%

\global\long\def\ESPERANZA#1{\mathbb{E}\PARENTESIS{#1}}%

\global\long\def\VARIANZA#1{\mathbb{V}\PARENTESIS{#1}}%

\global\long\def\COVARIANZA#1#2{\text{cov}\PARENTESIS{#1,#2}}%

\global\long\def\SINC{\text{sinc}}%


\global\long\def\POISSON#1{\text{Poisson}\PARENTESIS{#1}}%

\global\long\def\DIRACDELTA#1#2{\delta^{#2}\PARENTESIS{#1}}%
 

\selectlanguage{english}%

\qquad\qquad\qquad FERMILAB-PUB-20-620-E-QIS-T

\title{Ghost Imaging of Dark Particles  
}

\author{J.~Estrada\,$^{\pumpkin}$}
\affiliation{Fermi National Accelerator Laboratory, PO Box 500, Batavia IL, 60510, USA}
\author{R.~Harnik\,$^{\mathrightghost}$
}
\affiliation{Fermi National Accelerator Laboratory, PO Box 500, Batavia IL, 60510, USA}
\author{D.~Rodrigues\,$^{\mathrightbat}$
}
\affiliation{Department of Physics, FCEN, University of Buenos Aires and IFIBA, CONICET, Buenos Aires, Argentina}
\affiliation{Fermi National Accelerator Laboratory, PO Box 500, Batavia IL, 60510, USA}
\author{M.~Senger\,$^{\mathwitch*}$
}
\affiliation{Department of Physics, FCEN, University of Buenos Aires and IFIBA, CONICET, Buenos Aires, Argentina}
\affiliation{Physik-Institut der Universit\"at Z\"urich}

\date{October 31, 2020}

\begin{abstract}
We propose a new way to use optical tools from  quantum imaging and quantum communication to search for physics beyond the standard model. Spontaneous parametric down conversion (SPDC) is a commonly used source of entangled photons in which pump photons convert to a signal-idler pair.
We propose to search for ``dark SPDC'' (dSPDC) events in which a new dark sector particle replaces the idler. Though it does not interact,
the presence of a dark particle can be inferred by the properties of the  signal photon. Examples of dark states include axion-like-particles and dark photons. We show that the presence of an optical medium opens the phase space of the down-conversion process, or decay, which would be forbidden in vacuum. Search schemes are proposed which employ optical imaging and/or spectroscopy of the signal photons. 
The signal rates in our proposal scales with the second power of the feeble coupling to new physics, as opposed to light-shining-through-wall experiments whose signal scales with coupling to the fourth. We analyze the characteristics of optical media needed to enhance dSPDC and estimate the rate.  A bench-top demonstration of a high resolution ghost imaging measurement is performed employing a Skipper-CCD to demonstrate its utility in a dSPDC search. 
\end{abstract}

\maketitle


\section{Introduction}

Nonlinear optics is a powerful new tool for quantum information science. Among its many uses, it plays an enabling role in the areas of quantum networks and teleportation of quantum states as well as in quantum imaging.\blfootnote{$^{\pumpkin}$\,estrada@fnal.gov}
\blfootnote{$^{\mathrightghost}\,$roni@fnal.gov}
\blfootnote{$^{\mathrightbat}$\,rodriguesfm@df.uba.ar}
In quantum teleportation~\cite{furusawa_unconditional_1998,boschi_experimental_1998,bouwmeester_experimental_1997,bennett_teleporting_1993} the state of a distant quantum system, Alice, can be inferred without directly interacting with it, but rather by allowing it to interact with one of the photons in an entangled pair. The coherence of these optical systems has recently allowed teleportation over a large distance~\cite{2020arXiv200711157V}.
Quantum ghost imaging, or ``interaction-free'' imaging~\cite{white_interaction-free_1998}, is used to discern (usually classical) information about an object without direct interaction.
This technique exploits the relationship among the emission angles of a correlated photon pair to create an image with high angular resolution without placing the subject Alice in front of a high resolution detector or allowing it to interact with intense light.\blfootnote{$^{\mathwitch*}$\,matias.senger@physik.uzh.ch}
These methods of teleportation and imaging rely on the production of signal photons in association with an idler pair which is entangled (or at least correlated) in its direction, frequency, and sometimes polarization.
   
Quantum ghost imaging and teleportation both differ parametrically from standard forms of information transfer. This is because a system is probed, not by sending information to it and receiving information back, but rather by sending it half of an EPR pair, without need for a ``response''. The difference is particularly apparent if Alice is an extremely weekly coupled system, say she is part of a dark sector, with a coupling $\epsilon$ to photons. The rate of information flow in capturing an image of Alice will occur at a rate $\propto \epsilon^4$ with standard methods, but at a rate $\propto \epsilon^2$ using quantum optical methods. 

A common method for generating entangled photon pairs is the nonlinear optics process known as \emph{spontaneous parametric down conversion} (SPDC). 
In SPDC a pump photon decays, or down-converts, within a nonlinear optical medium into two other photons, a signal and an idler. 
The presence of the SPDC idler can be inferred by the detection of the signal~\cite{hong_mandel_1985}.  SPDC is in wide use in quantum information and provides the seed for  cavity enhanced setups such as optical parametric oscillators (OPO)~\cite{giordmaine_tunable_1965}.

\begin{figure}
\begin{centering}
\includegraphics[width=7cm]{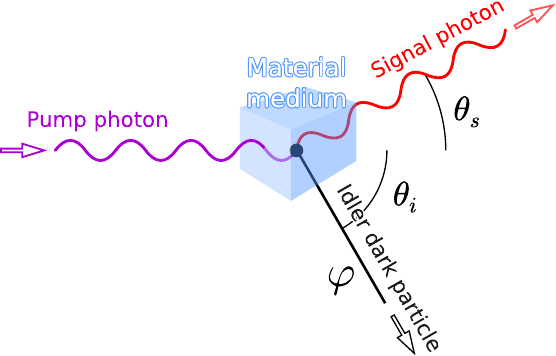}
\par\end{centering}
\caption{Pictorial representation of the dSPDC process. 
A dark particle $\varphi $ is emitted in association with a signal photon. The presence of $\varphi$ can be inferred from the distribution of the signal photon in angle and/or frequency. We consider both the colinear ($\theta_s=0$) and non-colinear ($\theta_s\ne 0$) cases. \label{fig:Pictorial-representation}}
\end{figure}
In this work we propose to use quantum imaging and quantum communication tools to perform an interaction-free search for the emission of new particles beyond the standard model.
The new tool we present is \emph{dark SPDC}, or dSPDC, an example sketch of which is shown in Figure~\ref{fig:Pictorial-representation}.
A pump photon enters an optical medium and down-converts to a signal photon and a dark particle, which can have a small mass, and does not interact with the optical medium. Like SPDC, in dSPDC the presence of a dark state can be inferred by the angle and frequency distribution of the signal photon that was produced in association. The dSPDC process can occur either collinearly, with $\theta_s=\theta_i=0$, or as shown in the sketch, in a non-colinear way.

The new dark particles which we propose to search for, low in mass and feebly interacting, are simple extensions of the standard model (SM). Among the most well motivated are axions, or axion-like-particles, and dark photons (see~\cite{essig_dark_2013} for a review). Since these particles interact with SM photons, optical experiments provide an opportunity to search for them. A canonical setup is light-shining-through-wall (LSW) which set interesting  limits on axions and dark photons using optical cavities~\cite{Hoogeveen:1990vq}. In optical LSW experiments a laser photon converts to an axion or dark photon, enabling it to penetrate an opaque barrier. For detection, however, the dark state must convert back to a photon and interact, which is a rare occurrence. As a result, the signal rate in LSW scales as $\epsilon^4$, where $\epsilon$ is a small coupling to dark states.  This motivates the interaction-free approach of dSPDC, in which the dark state is produced but does not interact again, yielding a rate $\propto\epsilon^2$.

From the perspective of particle physics, dSPDC, and also SPDC, may sound unusual\footnote{Searches for missing energy and momentum are a commonplace tool in the search for new physics at colliders~\cite{Goodman:2010yf,Bai:2010hh,Goodman:2010ku,Fox:2011pm,Aaltonen:2012jb}. However, the dSPDC kinematics are distinct as we show.}. In dSPDC, for example, a massless photon is decaying to another massless photon plus a massive particle. In the language of particle physics this would be a kinematically forbidden transition. The process, if it were to happen in vacuum, violates energy and momentum conservation and thus has no available phase space. %
In this paper we  show how optics enables us to perform ``engineering in phase space'' and open kinematics to otherwise forbidden channels. This, in turn, allows to design setups in which the dSPDC process is allowed and can be used to search for dark particles. 

Broadly speaking, the search strategies we propose may be classified as employing either imaging or spectroscopic tools, though methods employing both of these are also possible. In the first, the angular distribution of signal photons are measured, and in the second, the energy distribution is observed. For an imaging based search, a high angular resolution is required, while the spectroscopic approach requires high frequency resolution. The later has the benefit that dSPDC can be implemented in a waveguide which will enhance its rate for long optical elements~\cite{fiorentino_spontaneous_2007,ourpaper}.

We present the basic ideas and formalism behind this method, and focus on the phase space for dSPDC. We compare the phase space distributions of SPDC to dSPDC and discuss factors which may enhance the rate of dSPDC processes. To estimate the dSPDC rates in this work we re-scale known SPDC results. 
In a companion paper~\cite{ourpaper} we will derive the dSPDC rate more carefully, focusing on the spectroscopic approach and on \emph{colinear} dSPDC in bulk crystals and waveguides. 

In this work we also present a proof-of-concept experiment using a quantum imaging setup with a Skipper CCD~\cite{tiffenberg_single-electron_2017}. The high resolution is shown to produce a high resolution and low noise angular image of an SPDC pattern. Such a detector may be employed both in imaging and spectroscopic dSPDC setups.


This paper is structured as follows. In Section~\ref{sec:Lagrangian} we review axion-like particles and dark photons and present their Hamiltonians in the manner they are usually treated in nonlinear optics. 
In Section~\ref{sec:phase-matching} we discuss energy and momentum conservation, which are called \emph{phase matching conditions} in nonlinear optics, and the phase space for dSPDC.
The usual conservation rules will hold exactly in a thought experiment of an infinite optical medium. In Section~\ref{sec:finite} we discuss how the finite size of the optical medium broadens the phase space distribution and plot it as a function of the signal photon angle and energy.
In Section~\ref{sec:optimize} we discuss the choice of optical materials that allow for the dSPDC phase matching to enhance its rates and to suppress or eliminate SPDC backgrounds. In Section~\ref{sec:rates} we discuss the rates for dSPDC and its scaling with the geometry of the experiment, showing potential promising results for collinear waveguide setups with long optical elements.
In Section~\ref{sec:skipper}, we
present an experimental proof-of-concept in which a Skipper CCD is used to image an SPDC pattern with
high resolution.
We discuss future directions and conclude in Section~\ref{sec:conclusions}.

\section{The Dark SPDC Hamiltonian \label{sec:Lagrangian}}

The SPDC process can be derived from an effective nonlinear optics Hamiltonian of the form~\cite{ling_absolute_2008, Schneeloch2016}
\begin{equation}
    \mathcal{H}_\mathrm{SPDC} = \chi^{(2)} E_i E_s E_p
\end{equation}
where $E_j$ is the electric field for pump, signal, and idler photons and $j=p,s,i$ respectively. We adopt the standard notation in nonlinear optics literature in which the vector nature of the field is implicit. Thus the pump, signal and idler can each be of a particular polarization and $\chi^{(2)}$ represents the coupling between the corresponding choices of polarization.

We now consider the dark SPDC Hamiltonian 
\begin{equation}
\mathcal{H}_\mathrm{dSPDC}= g_{\varphi}\,\varphi E_s E_p 
\label{eq: term of the lagrangian}
\end{equation}
in which a pump and signal photon couple to a new field $\varphi$ which has a mass $m_\varphi$. The effective coupling $g_{\varphi}$, depends on the model and can depend on the frequencies in the process, the polarizations and the directions of outgoing particles.
This type of term arises
in well motivated dark sector models including axion-like particles and dark photons.
\subsection{Axion-like particles} 
Axion-like particles, also known as ALPs, are light pesudoscalar particles that couple to photons via a term 
\begin{equation}
    \mathcal{H}_\mathrm{axion}= 
    g_{\varphi}\, \varphi \vec E\cdot \vec B\,,
\end{equation}
which can be written in the scalar form of Equation~(\ref{eq: term of the lagrangian}) with the pump and signal photons chosen to have  orthogonal polarization. The coupling $g_\varphi$ in this case is the usual axion-photon coupling~$\sim 1/f_a$. ALPs are naturally light thanks to a spontaneously broken global symmetry at a scale $f_a$.

Since we will not perform a detailed analysis of backgrounds and feasibility here, we will describe existing limits on ALPs qualitatively and refer readers to limit plots in~\cite{Zyla:2020zbs,essig_dark_2013}.
In this work we will consider searches that can probe ALPs with a mass of order 0.1~eV or lower. Existing limits in this mass range come from stellar and supernova dynamics\cite{Raffelt:1990yz}, as well as from the CAST Heiloscope~\cite{ Anastassopoulos:2017ftl}, which require~$g_\varphi \lesssim 10^{-10}~\mathrm{GeV}^{-1}$. For comparison with our setups, we will also consider limits that are entirely lab-based, which are much weaker. At $\sim 0.1$~eV ALP coupling of order $10^{-6}~\mathrm{GeV}^{-1}$ are allowed by all lab-based searches. For axion masses around $10^{-3}-10^{-2}$~eV the PVLAS search\cite{DellaValle:2015xxa} for magnetic birefringence places a limit of $g_\varphi \lesssim 10^{-7}-10^{-6}~\mathrm{GeV}^{-1}$. Bellow masses of an meV, LSW experiments such as OSQAR~\cite{Ballou:2015cka} and ALPS~\cite{Ehret:2009sq} set $g_\varphi \lesssim 6\times 10^{-8}$. A large scale LSW facility, ALPS II~\cite{Bahre:2013ywa}, is proposed in order to improve upon astrophysical limits and CAST.

\subsection{Dark photons} A new U(1) gauge field, $A'$, with mass $m_{A'}$ that mixes kinetically with the photon via the Hamiltonian
\begin{equation}
    \mathcal{H}_\mathrm{mix} = \epsilon F^{\mu\nu}F'_{\mu\nu} = \epsilon \left(
    \vec E\cdot \vec E' + \vec B\cdot \vec B' \right)\,.
\end{equation}
It is possible to write the dark photon interaction in a basis  in which any electromagnetic current couples to longitudinally polarized dark photons with an effective coupling of $\epsilon (m_{A'}/\omega_{A'})$ and to transversely polarized dark photons with $\epsilon (m_{A'}/\omega_{A'})^2$~\cite{An:2013yfc,An:2013yua}. 
The same dynamics that leads to the nonlinear Hamiltonian $\mathcal{H}_\mathrm{SPDC}$, will yield 
 an effective coupling of two photons to a dark photon as in Equation~(\ref{eq: term of the lagrangian}) with $\varphi$ representing a dark photon polarization state. For longitudinal polarization of the dark photon this can be brought to the form of equation~\eqref{eq: term of the lagrangian} with $g_{\varphi} = \epsilon \chi^{(2)}  m_\varphi$. 
 The coupling to transverse dark photons will be suppressed by an additional factor of $m_\varphi/\omega_\varphi$. 

Like axions, limits on dark photons come from astrophysical systems, such as energy loss in the Sun~\cite{An:2013yfc}, as well as the lack of a detection of dark photons emitted by the Sun in dark matter direct detection experiments~\cite{An:2013yua}. At a dark photon mass of order 0.1~eV, the limit on the kinetic mixing is $\epsilon\lesssim 10^{-10}$ with the limit becoming weaker linearly as the dark photon mass decreases. Among purely lab-based experiments, ALPS sets a limit of $\epsilon\lesssim 3\times10^{-7}$. In forthcoming sections of this work we will use these limits as benchmarks.
\begin{center}
    \tiny{$\mathghost\mathghost\mathghost\mathghost$}
\end{center}
In addition to ALPs and dark photons, one can consider other new dark particles that are sufficiently light and couples to photons. Examples include millicharged particle, which can be produced in pairs in association with a signal photon. We leave these generalizations for future work and proceed to discuss the phase space for dSPDC in a model-independent way.


\section{Phase space of Dark SPDC}\label{sec:phase-matching}

As stated above, the effective energy and momentum conservation rules in nonlinear optics are known as \textquotedblleft phase matching conditions\textquotedblright. We begin by reviewing their origin to make the connection with energy and momentum conservation in particle physics and then proceed to solve them for SPDC and dSPDC, to understand the phase space for these processes.

\subsection{Phase matching and particle decay}
We begin with a discussion of the phase space in a two-body decay, or spontaneous down-conversion. We use particle physics language, but will label the particles as is usual in the nonlinear optics systems which we will be discussing from the onset. 
A pump particle $p$ decays into a signal particle $s$ and another particle, either an idler photon $i$ or a dark particle $\varphi$:
\begin{eqnarray}
\mbox{SPDC:}&\qquad&
\gamma_{p}\to\gamma_{s}+\gamma_{i}\nonumber\\
\mbox{dSPDC:}&\qquad& \gamma_{p}\to\gamma_{s}+\varphi\,.
\label{eq: process equation}
\end{eqnarray}
We would like to discuss the kinematics of the standard model SPDC process, and the BSM dSPDC process together, to compare and contrast. For this, in the discussion below we will use the idler label $i$ to represent the idler photon in the SPDC case, and $\varphi$ in the dSPDC case. Hence $\VECTOR{k}_i$ and $\theta_i$ are the momentum and emission angle of either the idler photon or of $\varphi$, depending on the process.

The differential decay rate in the laboratory frame is given by~\cite{Zyla:2020zbs}
\begin{equation}\label{eq:dif-rate}
    \frac{d\Gamma}{d^3k_s d^3k_i} = 
    \frac{\delta^4\left(p_{p\mu}-p_{s\mu}-p_{i\mu}\right)}{(2\pi)^2\, 2\omega_p\, 2\omega_s\, 2\omega_i} |\mathcal{M}|^2
\end{equation}
where $\mathcal{M}$ is the amplitude for the process (which carries a mass dimension of~$+1$), and the energy-momentum four-vectors
\begin{equation}
    p_{j\mu} = \left(
    \begin{array}{c}
         \omega_j \\ \VECTOR k_j
    \end{array} \right)
\end{equation}
with $j=p,s,i$. Traditionally, one proceeds by integrating over four degrees of freedom within the six dimensional phase space, leaving a differential rate with respect to the two dimensional phase space. This is often chosen a solid angle for the decay $d\Omega$, but in our case there will be non-trivial correlations of frequency and angle as we will see in the next subsection.

It is instructive, however, for our purpose to take a step back and recall how
the energy-momentum conserving $\delta$-functions come about. 
In calculating the quantum amplitude for the transition,
 the fields in the initial and final states are expanded in modes. The plane-wave phases $e^{i\PARENTESIS{\omega t-\VECTOR k\cdot\VECTOR x}}$ are collected from each and a spacetime integral is performed 
\begin{equation}
\int d^4x\, e^{i\PARENTESIS{\Delta\omega t-\Delta\VECTOR k\cdot\VECTOR x}} = (2\pi)^4\, \delta^{(3)}(\Delta\VECTOR k)\,
\delta (\Delta\omega)\,, \label{eq:delta-func}
\end{equation}
where frequency and momentum mismatch are defined as
\begin{equation}
\LKEY{\begin{aligned} & \Delta\omega=\omega_{p}-\omega_{s}-\omega_{i}\\
 & \Delta\VECTOR k=\VECTOR k_{p}-\VECTOR k_{s}-\VECTOR k_{i}
\end{aligned}\text{,}
}\label{eq: definition of Delta omega and Delta k}
\end{equation}
and again, the label $i$ can describe either an idler photon or a dark particle $\varphi$.
In Equation~(\ref{eq: definition of Delta omega and Delta k}), we separated momentum and energy conserving delta functions to pay homage to the optical systems which will be the subject of upcoming discussion. The squared amplitude $|\mathcal{M}|^2$, and hence the rate, is proportional to a single power of the energy-momentum conserving delta function times a space-time volume factor, which is absorbed for canonically normalized states (see e.g.~\cite{TongNotes}), giving Equation~(\ref{eq:dif-rate}). 

The message from this is that energy and momentum conservation, which are a consequence of Neother's theorem and space-time translation symmetry, is enforced in quantum field theory by perfect destructive interference whenever there is a non-zero mismatch in the  momentum or energy of initial and final states. In this sense, the term ``phase matching'' captures the particle physicist's notion of energy and momentum conservation  well.   
 
The dSPDC process shown in Figure~\ref{fig:Pictorial-representation} is a massless pump photon decaying to a signal photon plus a massive particle $\varphi$. This process clearly cannot occur in vacuum. For example, if we go to the rest frame of $\varphi$, and define all quantities in this frame with a tilde $\tilde\,$, the conservation of momentum implies $\VECTOR{\tilde k}_p=\VECTOR{\tilde k}_s$, while the conservation of energy implies $\tilde \omega_p=\tilde \omega_s+m$. Combining these with the dispersion relation for a photon in vacuum $\tilde\omega_{p,s}=|\VECTOR{\tilde k}_{p,s}|$, implies that energy and momentum conservation cannot be satisfied and the process is kinematically forbidden. Another way to view this is to recall that the energy-momentum four-vector for the initial state (a photon) lies on a light cone i.e. it is a null four-vector, $p^\mu p_\mu = 0$, while the final state cannot accomplish this since there is a nonzero mass.
 
In this work we show that optical systems will allow us to open phase space for dSPDC. When a photon is inside an optical medium a different dispersion relation holds,
\begin{equation}
    \tilde n_p \tilde \omega_p= |\VECTOR{\tilde k}_{p}| \qquad\mbox{and}\qquad  \tilde n_s \tilde \omega_s= |\VECTOR{\tilde k_{s}}|
\end{equation}
where $\tilde n_p$ and $\tilde n_s$ are indices of refraction of photons in the medium in the $\varphi$ rest frame, which can be different for pump and signal. 
As we will see later on, under these conditions the conclusion that $p\to s+\varphi$ is kinematically forbidden can be evaded.

Another effect that occurs in optical systems, but not in decays in particle physics, is the breaking of spatial translation invariance by the finite extent of the optical medium. This allows for violation of momentum conservation along the directions in which the medium is finite. As a result, the sharp momentum conserving delta function will become a peaked distribution of width $L^{-1}$, where $L$ is the crystal size.
We will begin by solving the exact phase matching conditions in SPDC and dSPDC in the following subsection, which correspond to the phase space distributions for an infinitely large crystal. Next, we will move on to the case in which the optical medium is finite.

\subsection{Phase Matching in dSPDC}

We now study the phase space for dSPDC to identify the correlations between the emission angle and the frequency of the signal photon. We will consider in parallel the SPDC process as well, so in the end we arrive to both results. We assume in this subsection an optical medium with an infinite extent, such that the delta-functions enforce energy and momentum conservation.

The phase matching conditions $\Delta \omega=0$ and $\Delta\VECTOR k = 0$ ---that must be strictly satisfied in the infinite extent optical medium scenario--- imply
\begin{equation}
\LKEY{\begin{aligned} & \omega_{p}=\omega_{s}+\omega_{i}\\
 & \VECTOR k_{p}=\VECTOR k_{s}+\VECTOR k_{i}
\end{aligned}\,.
}\label{eq: dsaojdbjdspdasq}
\end{equation}
Using the $\VECTOR k$'s decomposition shown in Figure~\ref{fig:Pictorial-representation} the second equation can be expanded in coordinates parallel and perpendicular to the pump propagation direction 
\begin{equation}
\LKEY{\begin{aligned} & k_{p}=k_{s}\cos\theta_{s}+k_{i}\cos\theta_{i}\\
 & 0=k_{s}\sin\theta_{s}-k_{i}\sin\theta_{i}
\end{aligned}
}\label{eq: dsjiadsaobdois}
\end{equation}
 where the angles $\theta_{s}$ and $\theta_{i}$ are those indicated
in the cited figure. Since this process
is happening in a material medium, the photon dispersion relation is $k=n\omega$
 where $n$ is the refractive index. The refractive index is in general a function of frequency, polarization, and the direction of propagation, and can thus be different for pump, signal, and idler photons.
For dSPDC, in which the $\varphi$ particle
is massive and weekly interacting with matter the dispersion relation is the same
as in vacuum,
\begin{equation}
k_\varphi=\sqrt{\omega_\varphi^{2}-m_\varphi^{2}}\text{.}
\end{equation}
 Thus we can explicitly write the dispersion relation for each particle
in this process: 
\begin{eqnarray}
k_{p}=n_{p}\omega_{p}\text{,}\label{eq: kp igual np omegap}\\
k_{s}=n_{s}\omega_{s}\label{eq: ks igual ns omegas}
\end{eqnarray}
 and 
\begin{equation}
k_{i}=\LKEY{\begin{aligned} & n_{i}\omega_{i} &  & \text{for SPDC}\\
 & \sqrt{\omega_{i}^{2}-m_\varphi^{2}} &  & \text{for dSPDC}
\end{aligned}\text{.}
}\label{eq: ki igual cosa rara}
\end{equation}
where, again, the label ``$i$'' refers to an idler photon for SPDC and to the dark field $\varphi$ for dSPDC.
Note that $k_{i}$ for SPDC
not only has $m=0$ but also takes into account the
``strong (electromagnetic) coupling'' with matter summarized by
the refractive index~$n_{i}$. 
As a result, for SPDC we always have $k_i>\omega_i$ while for dSPDC $k_i<\omega_i$.
Replacing these dispersion relations
into~(\ref{eq: dsjiadsaobdois}) and using the first equation in~(\ref{eq: dsaojdbjdspdasq})
we obtain the phase matching relation of signal angle and frequency
\begin{equation}
\cos\theta_{s}=\frac{n_{p}^{2}+\alpha_\omega^{2}n_{s}^{2}-\Xi^{2}}{2\alpha_\omega n_{p}n_{s}}\label{eq: cos theta signal phase matching}
\end{equation}
 where we define 
\begin{equation}
\alpha_\omega\DEF\frac{\omega_{s}}{\omega_{p}}\label{eq: definition of alpha}
\end{equation}
 and 
\begin{equation}
\Xi\DEF
\frac{k_i}{\omega_p} =
\LKEY{\begin{aligned} & n_{i}\PARENTESIS{1-\alpha_\omega} &  & \text{for SPDC}\\
 & \sqrt{\PARENTESIS{1-\alpha_\omega}^{2}-\frac{m_\varphi^{2}}{\omega_{p}^{2}}} &  & \text{for dSPDC}
\end{aligned}
}\text{.}\label{eq:def-Xi}
\end{equation}
Equation~(\ref{eq: cos theta signal phase matching}) defines the phase space for (d)SPDC, along with the azimuthal angle $\phi_s$.
\begin{figure*}
    \centering
    \includegraphics[width=0.9\columnwidth]{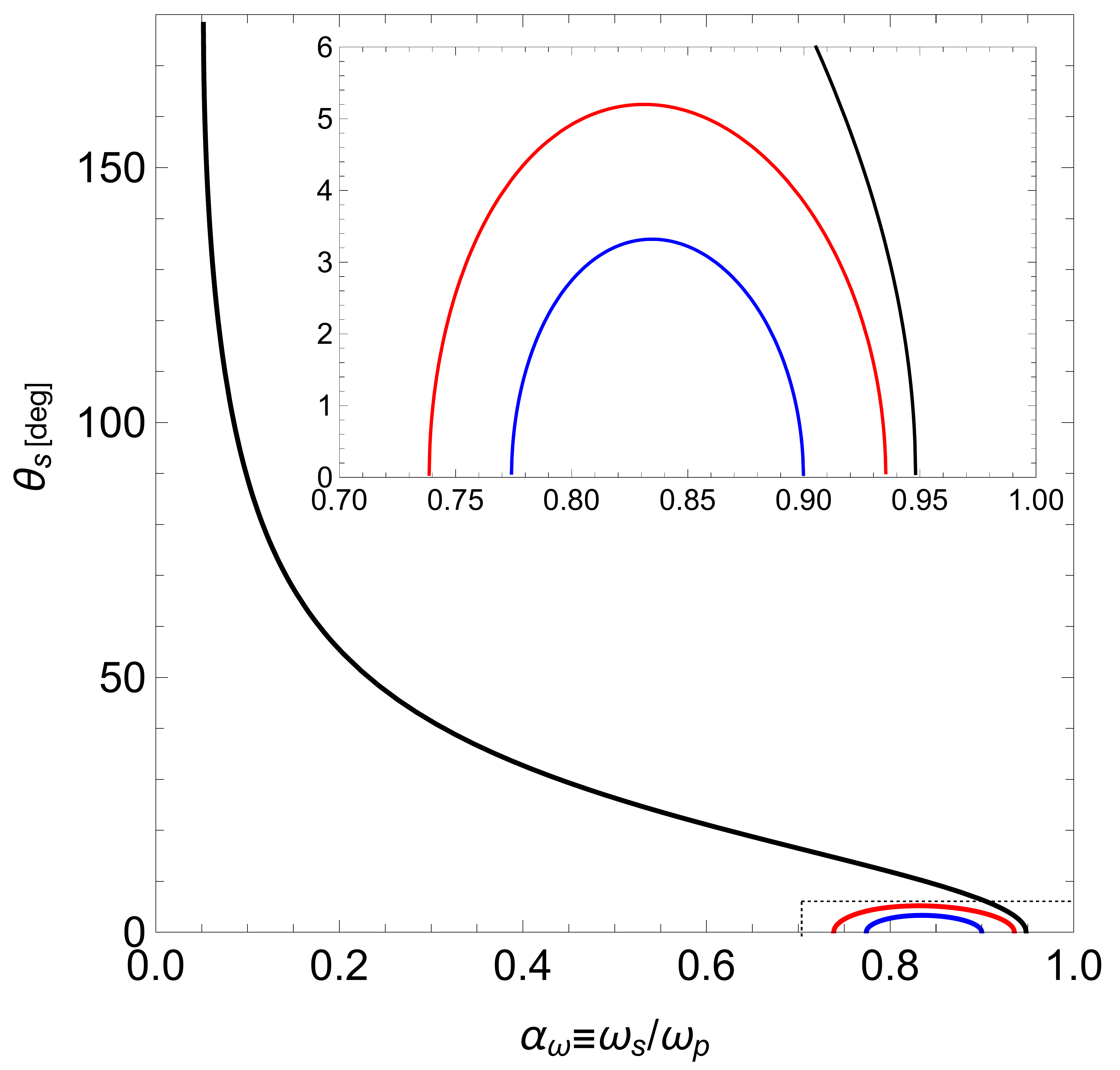}
  \qquad
    \includegraphics[width=\columnwidth]{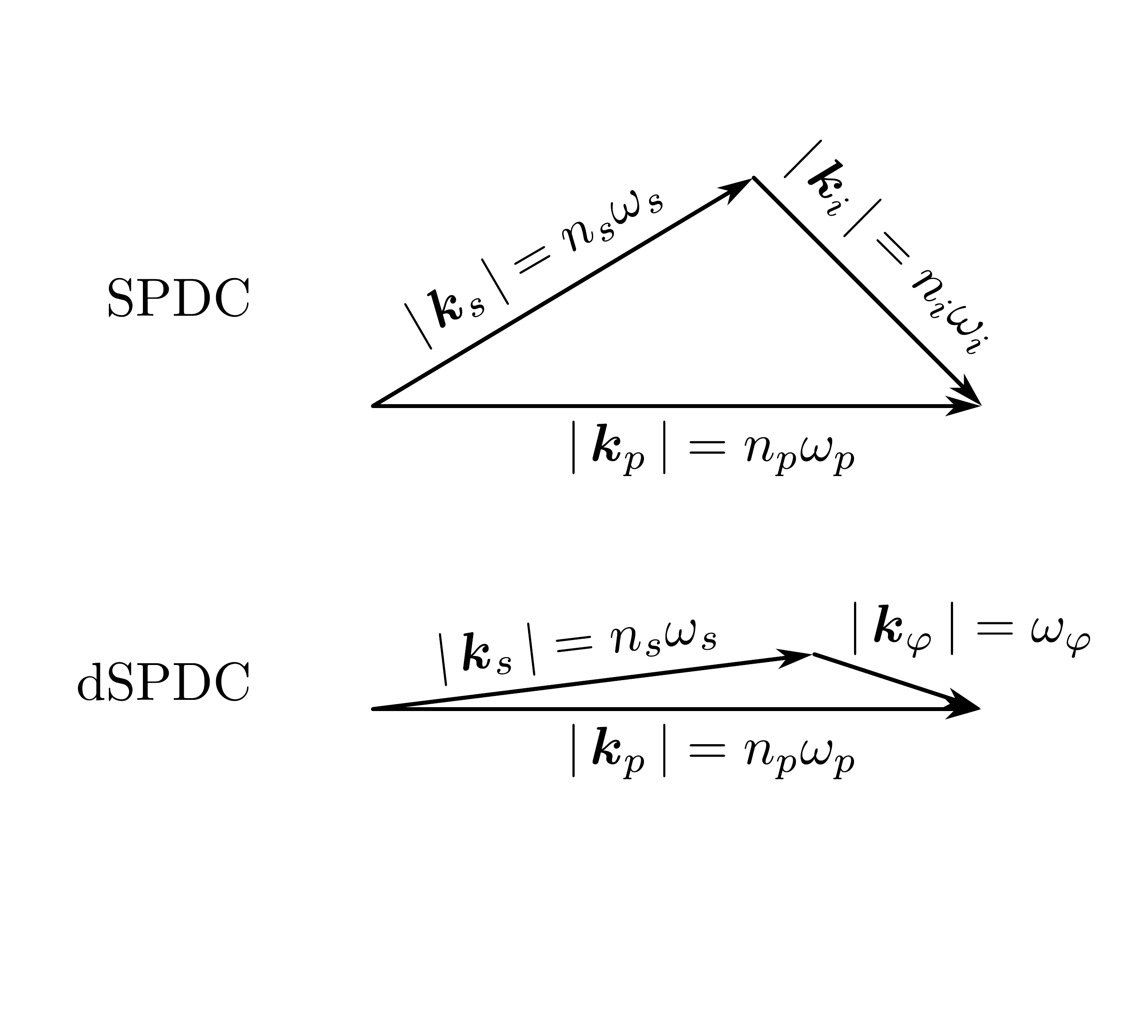}
    \caption{\emph{Left:} The allowed phase space for SPDC (black), dSPDC with $m=0$ (red), and dSPDC with $m=0.1\omega_p$ (blue) shown as in the plane of signal emission angle as a function of frequency ratio $\alpha_\omega$. The indices of refraction here are $n_p=1.658$ and $n_s=1.486$ as an example, inspired by  calcite. The inset shows a zoom-in of the dSPDC phase space. \emph{Right:} Sketches depicting the momentum phase matching condition $\Delta \VECTOR k = 0$ for SPDC and dSPDC, with massless $\varphi$. In both cases we take the same $\omega_p$ and $\omega_s$. Due to the index of refraction for $\varphi$ is essentially one, and that for the idler photon is larger (say $\sim 1.5$), phase matching in dSPDC has a smaller signal emission angle than that of SPDC. 
    }
    \label{fig:phasespace}
\end{figure*}
The idler angle, or that of the dark particle in the dSPDC case, is fixed in terms of the signal angle and frequency by requiring conservation of transverse momentum
\begin{eqnarray}
    \sin \theta_i &=& \frac{k_s}{k_i}\sin\theta_s\nonumber\\
    \phi_i &=& \phi_s+\pi \label{eq:k_iT}
\end{eqnarray}
where $k_i$ is evaluated at a frequency of $\omega_i = \omega_p-\omega_s$, and $k_s$ at the frequency $\omega_s$ according to the dispersion relation in 
Equations~(\ref{eq: ks igual ns omegas}) and~(\ref{eq: ki igual cosa rara}). 

In the left panel of Figure~\ref{fig:phasespace} we show the allowed phase space in the $\alpha_\omega$-$\theta_s$ plane for SPDC and for dSPDC with both a massless and massive $\varphi$. Here we have chosen $n_p=1.486$, $n_s=1.658$ with $n_i=n_s$ in the SPDC case. These values were chosen to be constant with frequency and propagation direction, for simplicity and are inspired by the ordinary and extraordinary refractive indices in calcite and will be used as a benchmark in some of the examples below.

We see that phase matching is achieved in different regions of phase space for SPDC and its dark counterpart. In dSPDC, signal emission angles are restricted to near the forward region and in a limited range of signal frequencies. The need for more forward emission in dSPDC can be understood because $\varphi$ effectively sees an index of refraction of 1. 
This implies it can carry less momentum for a given frequency, as compared to a photon which obeys $k=n\omega$. As a result, the signal photon must point nearly parallel to the pump, in order to conserve momentum (see Figure~\ref{fig:phasespace}, right).
 An additional difference to notice is that, in contrast with the SPDC example shown, for a fixed signal emission angle there are two different signal frequencies that satisfy the phase matching in dSPDC. This will always be the case for dSPDC. Although SPDC can be of this form as well, typical refractive indices usually favor single solutions as shown for calcite.

\subsubsection{Phase matching for colinear dSPDC}
One case of particular interest is that of colinear dSPDC, in which the emission angle is zero, as would occur in a single-mode fiber or a waveguide. 
For this case, so long as $n_s>n_p$, an axion  mass below some threshold can be probed. 
 \begin{figure}
     \centering
     \includegraphics[width=0.9\columnwidth]{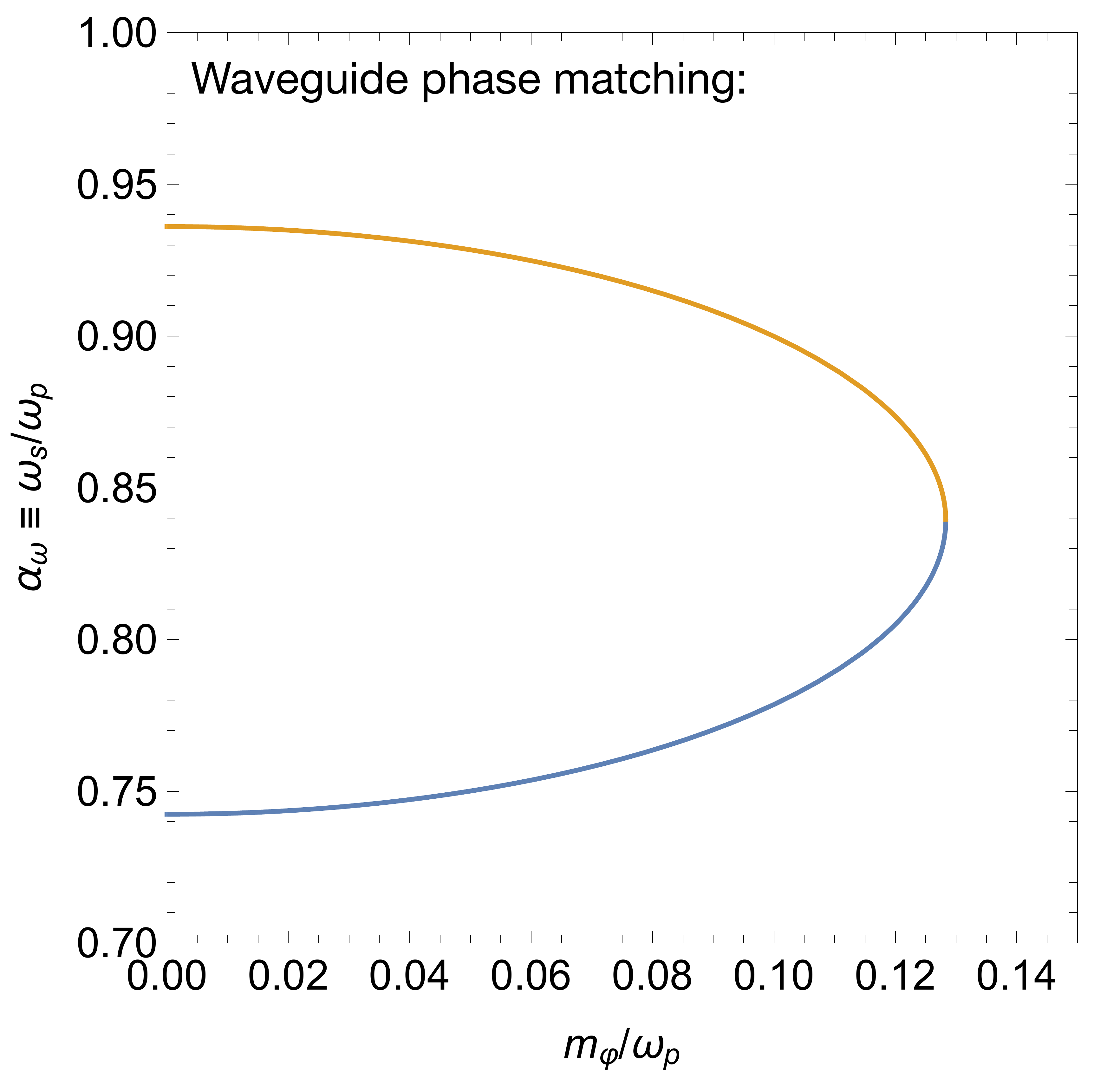}
     \caption{Solutions to the phase matching conditions for the colinear dark SPDC process for the signal photon energy as a function of the dark particle $\varphi$ mass. This phase space is relevant for waveguide-based experiments.
     Both axes are normalized to the pump frequency. The two branches correspond to configurations in which $\varphi$ is emitted forward (bottom) and backward (top).}
     \label{fig:phase-match-calcite}
 \end{figure}
Setting the emission angles to zero there are two solutions to the phase matching equations which give signal photon energies of
\begin{equation}
\alpha_\omega = \frac{\left(n_sn_p -1\right) \pm \sqrt{\left(n_s-n_p \right)^2 - \left(n_s^2-1\right)\frac{m_\varphi^2}{\omega_p^2}}}{n_s^2-1}\,.
\end{equation}
The $+$ solution above corresponds to the case where the $\varphi$ particle is emitted in the forward direction whereas the $-$ represents the case where it is emitted backwards. The signal frequency as a function of $\varphi$ mass is shown in Figure~\ref{fig:phase-match-calcite} 
for the calcite benchmark.
For the near-massless case, $m_{\varphi}\ll\omega_p$, a phase matching  solution exists for $n_s\ge n_p$, giving
\begin{equation}
    \frac{\omega_s}{\omega_p} = \frac{n_p\mp 1}{n_s\mp 1}\,.
\end{equation}
For calcite, we get~$\omega_s=0.739 \omega_p$ and~$\omega_s=0.935 \omega_p$ for forward and backward emitted axions respectively.

Of course, not any combination of $n_p$ and $n_s$ will allow to achieve phase matching in dSPDC.
We will state here that $n_s>n_p$ is a requirement for phase matching to be possible for a massless $\varphi$ and that $n_s-n_p$ must grow as $m_\varphi$ grows. 
We will discuss these and other requirements for dSPDC searches in Section~\ref{sec:optimize}.

\subsection{Thin planar layer of optical medium}\label{sec:finite}

We now consider the effects of a finite crystal.  Specifically, we will assume that the optical
medium is a planar thin layer\footnote{The meaning of what constitutes the thin crystal layer limit in our context will be discussed in Section~\ref{sec:rates}.} of optical material of length $L$ along the pump propagation direction $\hat z$, and is infinite in transverse directions.
In this case, the integral in Equation~(\ref{eq:delta-func}) is performed only over a finite range and the delta function is replaced by a $\SINC\ x \equiv \sin x / x$ function
\begin{equation}
\int_{-L/2}^{L/2} dz\, e^{i \Delta k_z z} = L\,\mathrm{sinc}\left(\frac{\Delta k_z L}{2}\right)
\end{equation}
at the level of the amplitude. 
As advertised, this allows for momentum non-conservation with a characteristic width of order $L^{-1}$. 

The fully differential rate 
will be proportional to the squared sinc function
\begin{eqnarray}
\label{eq:dif-rate-sinc}
    \frac{d\Gamma}{d^3 k_p d^3 k_i}
    &\propto& L^2\mathrm{sinc}^2 \left(\frac{\Delta k_zL}{2}\right)\delta^2(\Delta \VECTOR k_T)\delta(\Delta\omega)\\
    &\equiv& 
    \frac{d\hat\Gamma}{d^3 k_p d^3 k_i}\label{eq:deffinition of phase space distribution}
\end{eqnarray}
where $\Delta \VECTOR k_T$ is the momentum mismatch vector in the transverse directions. 
The constant of proportionality in Equation~(\ref{eq:dif-rate-sinc}) will have frequency normalization factors of the form $(2\omega)^{-1}$ and the matrix element $\mathcal{M}$. The matrix element can also have non trivial angular dependence in $\theta$ as well as in $\phi$, depending on the model and the optical medium properties.
Note, however, that the current analysis for the phase space is model independent. We thus postpone the discussion of overall rates to Section~\ref{sec:rates} and to~\cite{ourpaper} and here we limit the study to the phase space distribution only. We will proceed with the defined phase space distribution $d\hat\Gamma$ in Equation~(\ref{eq:deffinition of phase space distribution}), which we will now study.

One can trivially perform the integral over the $d^3k_i$ which will effectively enforce Equation~(\ref{eq:k_iT}) for transverse momentum conservation, and set $|k_{i}|$ by conservation of energy. The argument of the sinc function is
\begin{eqnarray}
\Delta k_z &=& k_p-k_s \cos\theta_s - k_i \cos\theta_i \\
&=& \omega_p\left( n_p - n_s \alpha_\omega \cos\theta_s \pm \sqrt{\Xi^2- n_s^2 \alpha_\omega^2\sin^2\theta_s} \right) \nonumber
\end{eqnarray}
where in the second step we have used Equation~(\ref{eq:k_iT}) and the definitions of $\alpha_\omega$ and $\Xi$ in Equations~(\ref{eq: definition of alpha}) and~(\ref{eq:def-Xi}). The $\pm$ accounts for the idler or dark particle being emitted in the forward and backward direction respectively. 

\begin{widetext}
It is convenient to re-express the remaining three-dimensional phase space for (d)SPDC in terms of the signal emission angles $\theta_s$ and $\phi_s$, and the signal frequency (or equivalently $\alpha_\omega$). Within our assumptions, the distribution in $\phi_s$ is flat. With respect to the polar angle and frequency we find 
\begin{equation}
\frac{d^2\hat\Gamma}{d (\cos \theta_s) d \alpha_\omega}=
\frac{2\pi \omega_p^3 \alpha_{\omega}^{2}\PARENTESIS{1-\alpha_{\omega}}n_{s}^{3}\tilde{n}_{i}^{2}
}{\sqrt{\Xi^{2}-\alpha_{\omega}^{2}n_{s}^{2}\sin^{2}\theta_{s}}}
\sum_{\pm}L^2\SINC^{2}\CORCHETES{\PARENTESIS{n_p - n_s \alpha_\omega \cos\theta_s \pm \sqrt{\Xi^2- n_s^2 \alpha_\omega^2\sin^2\theta_s}}\frac{\omega_p L}{2}}\label{eq: F_alpha_theta_phi}
\end{equation}
\end{widetext}
 where 
\begin{equation}
\tilde{n}_{i}\DEF\LKEY{\begin{aligned} & n_{i} &  & \text{for SPDC}\\
 & 1 &  & \text{for dSPDC}
\end{aligned}
}\text{.}
\end{equation}

\begin{figure}
\begin{center}
\includegraphics[width=0.9\columnwidth]{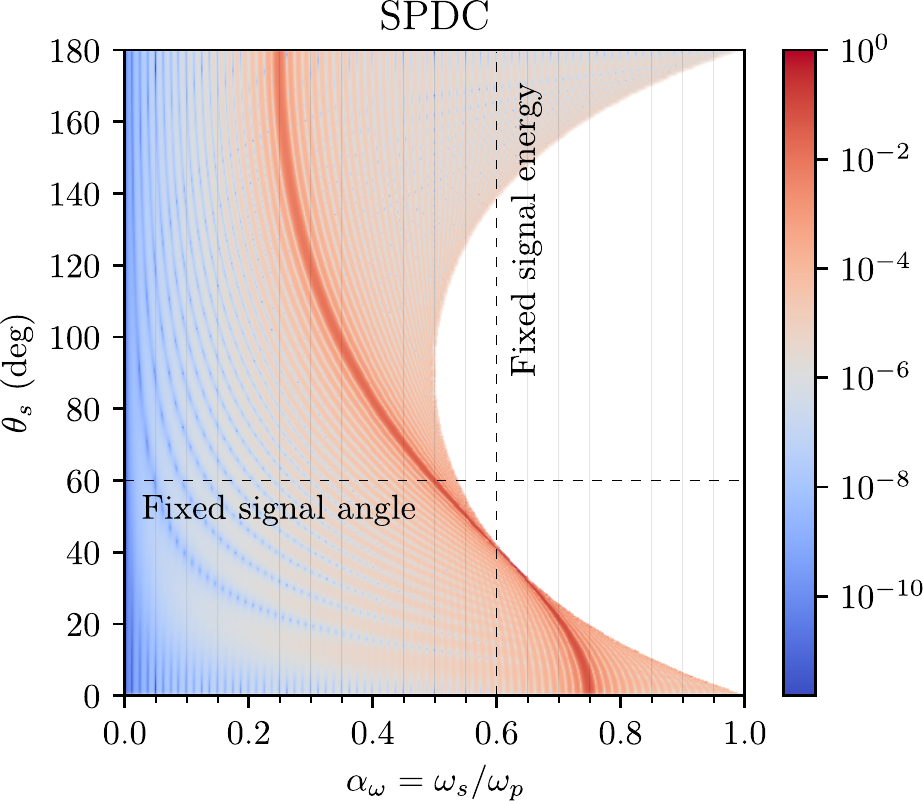}
\end{center}
\caption{The phase space distribution $d^2\hat\Gamma/d(\cos\theta_s) d\alpha_\omega$ for SPDC in arbitrary units. 
We use $L=1$, $\lambda_{p}=\frac{L}{10}$, $n_{p}=2$ and $n_{s}=n_{i}=4$. These parameters are unrealistic in practice, but allow for clear visualization of the distribution. Dashed lines show slices of fixed signal angle/energy. \label{fig: F_SPDC}}
\end{figure}

\begin{figure*}
\begin{centering}
\includegraphics[width=1\columnwidth]{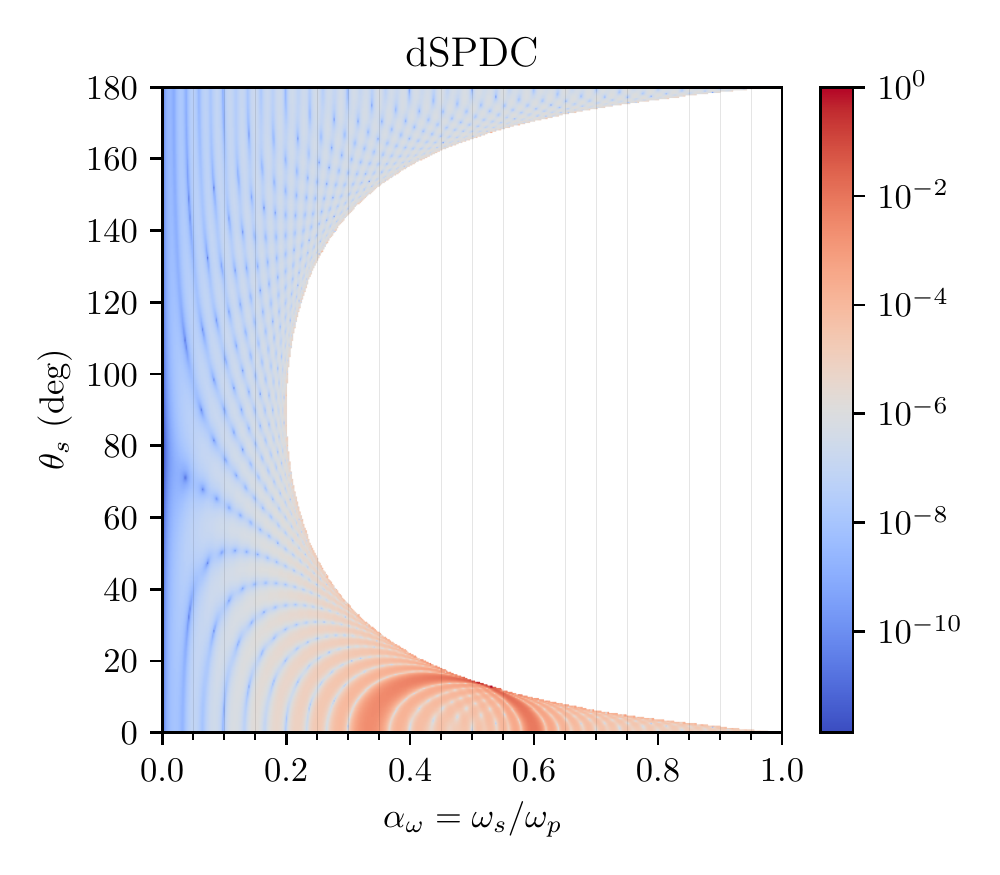}
\includegraphics[width=1\columnwidth]{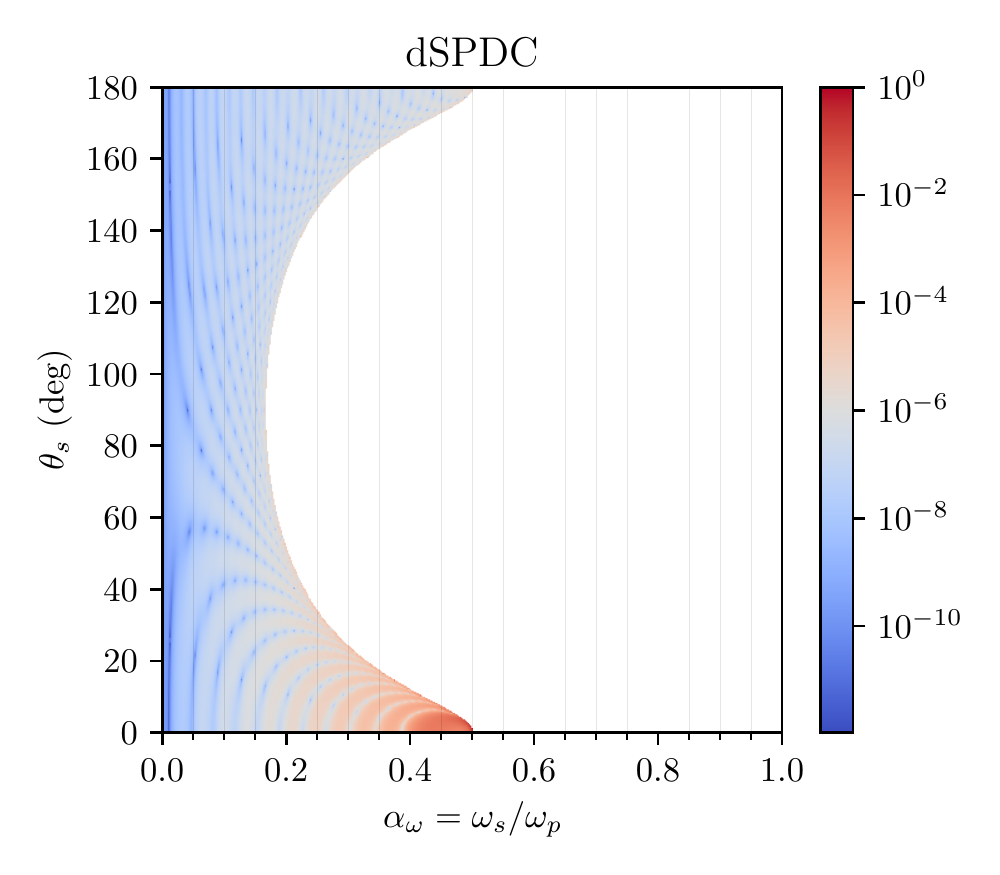}
\caption{Phase space distribution $d^2\hat\Gamma/d(\cos\theta_s) d\alpha_\omega$ in arbitrary units for the dSPDC process for $m=0$ (left) and $m=0.5 \omega_p$. The other parameters are as in Figure~\ref{fig: F_SPDC}, with somewhat exaggerated indices of refraction to allow seeing the features in the distribution. \label{fig: F_dSPDC}}
\end{centering}
\end{figure*}

In Figure~\ref{fig: F_SPDC} we show the double differential distribution $\hat\Gamma$ for SPDC, and in Figure~\ref{fig: F_dSPDC} it is shown for dSPDC both for a massless and a massive~$\varphi$. The distribution clearly peaks in regions of phase space that satisfy the phase matching conditions, those shown in Figure~\ref{fig:phasespace}. To have a clearer view of the qualitative features in the distribution we picked somewhat exaggerated values for indices of refraction in these figures.

\subsubsection{dSPDC Searches with Imaging or Spectroscopy}
Though one can, in principle, measure both the angle and the energy of the signal photon, it is usually easier to measure either with a fixed angle or a fixed energy. 
For example, a CCD detector with a monochromatic filter can easily measure a single energy slice of this distribution, as shown in the vertical dashed line in Figure~\ref{fig: F_SPDC}. Likewise, a spectrometer with no spatial resolution can measure a fixed angle slice of this distribution, as shown in the horizontal dashed line of Figure~\ref{fig: F_SPDC}. 
An attractive choice for this is to look in the forward region, with an emission angle $\theta_s=0$, which is the case for waveguides. 
For this colinear case we get a signal spectrum
\begin{equation}
    \frac{d\hat\Gamma}{d\alpha_\omega}=
    \frac{2\pi \omega_p^3 \alpha_{\omega}^{2}\PARENTESIS{1-\alpha_{\omega}}n_{s}^{3}\tilde{n}_{i}^{2}
}{\Xi}
\sum_\pm
L^2
\mathrm{sinc}^2\left(\frac{\Delta k_{z\pm}^{(0)}L}{2}\right)
\end{equation}
where 
\begin{equation}
    k_{z\pm}^{(0)} = k_{zp}-k_{zs}-k_{zi}^\pm =
    \omega_p\left( n_p - n_s \alpha_\omega \pm \Xi \right)
\end{equation}
is the momentum mismatch for colinear (d)SPDC. In some cases the colinear spectrum is dominated by just the forward emission of the idler/$\varphi$, while in others there are two phase matching solutions which contribute similarly to the rate. 

\begin{figure*}
\begin{centering}
\includegraphics[width=1\columnwidth]{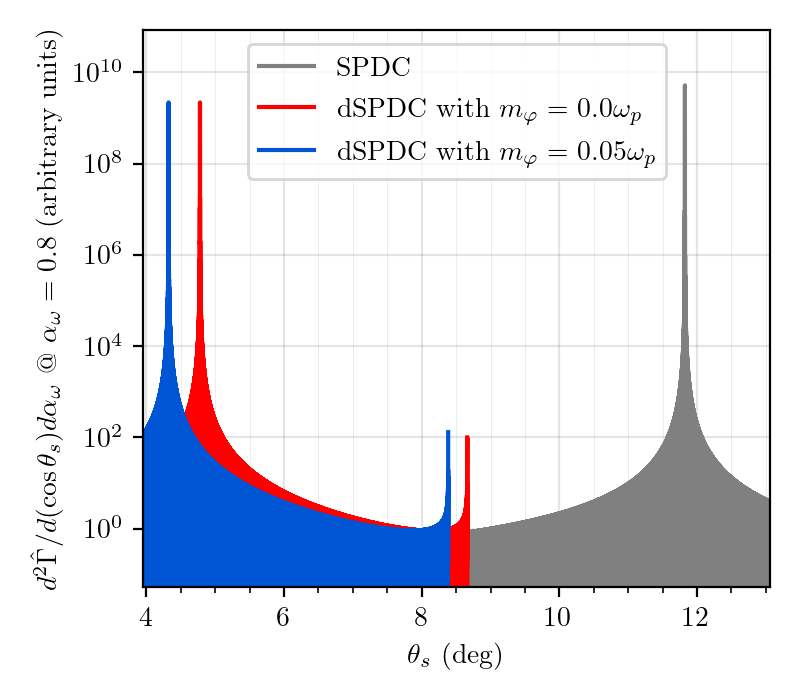}
\includegraphics[width=1\columnwidth]{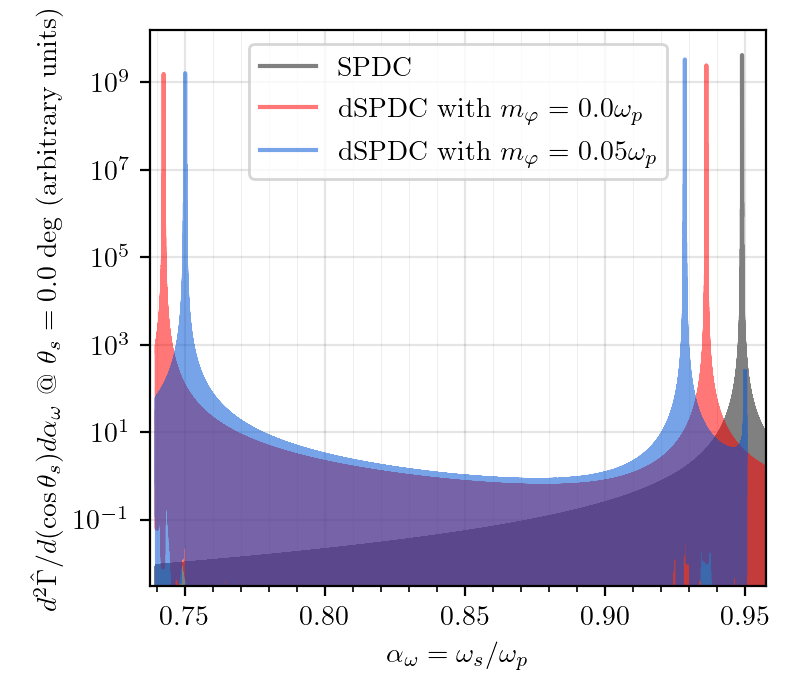}
\par\end{centering}
\caption{Slices of the phase space distributions for SPDC and for dSPDC with different choices of $\varphi$ mass. On the left we show angle distributions with fixed signal frequency, and on the right we show frequency distributions for fixed signal angle. These distributions allow to separate SPDC backgrounds  from the dSPDC signal, as well as the dSPDC signal for different values of $m_\varphi$. The distributions become narrower for thicker crystals, enhancing the signal to background separation power. For these distributions we used $L=1\text{ cm}$, $\lambda_p = 400 \text{ nm}$ and for the refractive indices $n_p = 1.49$ and $n_s = n_i = 1.66$ inspired by calcite.
\label{fig:energy and angle slices plots}}
\end{figure*}

In Figure~\ref{fig:energy and angle slices plots} we show two distributions, one for fixed signal frequency, and the other for a fixed signal angle, the latter in the forward direction. We see that both of these measurement schemes  are able to distinguish the signal produced by different values of mass $m_\varphi$. Furthermore, in cases where the standard model SPDC process is a source of background, it can be separated from the dSPDC signal. 
As expected, in the forward measurement the dSPDC spectrum exhibits two peaks for the emission of $\varphi$ in the forward or backward directions.
It should be noted that the width  
of the highest peaks in these distributions decreases with crystal length $L$.


\section{Optical Materials for  Dark SPDC}\label{sec:optimize}

Having discussed the phase space distribution for dSPDC, we will now discuss which optical media are needed to open this channel, to enhance its rates, and, if possible, to suppress the backgrounds.
A more complete estimate of the dSPDC rate will be presented in~\cite{ourpaper}. Here we will discuss the various features qualitatively to enable optimization of dSPDC searches. 

\subsection{Refractive Indices}\label{sec:indices}

As we discussed above, the refractive index of the optical medium plays a crucial role. It opens the phase space for the dSPDC decay and determines the kinematics of the process.

In order to enhance the dSPDC rate, it is always desirable to have a setup in which the dSPDC phase matching conditions which we discussed in Section~\ref{sec:phase-matching} are accomplished.   
Since the left hand side of Equation~(\ref{eq: cos theta signal phase matching})
is the cosine of an angle, its right hand is restricted between $\pm1$ so
\begin{equation}
-1<\frac{n_{p}^{2}+\alpha_\omega^{2}n_{s}^{2}-\Xi^{2}}{2\alpha_\omega n_{p}n_{s}}<1\text{.}\label{eq: condition for the phase matching conditions}
\end{equation}
From this one finds that so long as $n_s>n_p$, there is a range of $\varphi$ mass which can take part in dSPDC. 
As the difference $n_s-n_p$ is taken to be larger, a greater $\varphi$ mass can be produced and larger opening angles $\theta_s$ can be achieved. 

\begin{figure}
\begin{centering}
\includegraphics{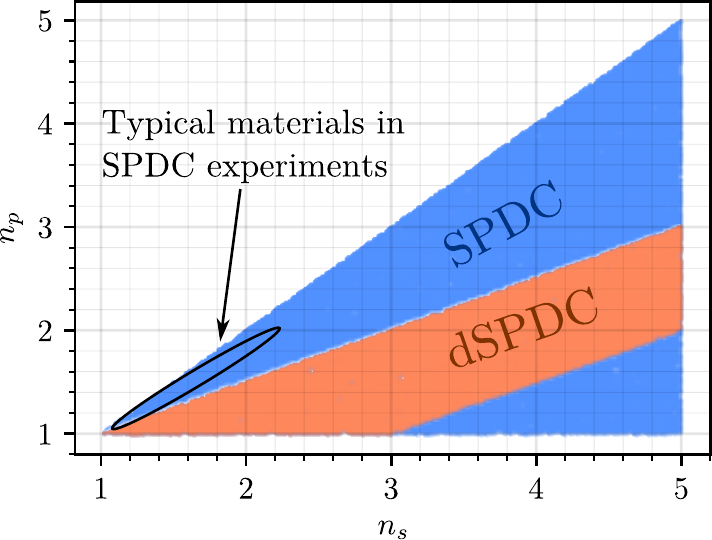}
\par\end{centering}
\caption{Regions in which the refractive indices $n_{p}$ and $n_{s}$ allow
the phase matching with $\alpha_\omega\equiv \omega_s/\omega_p=\frac{1}{2}$ for the SPDC process
(in this case assuming $n_{i}=n_{s}$) and dSPDC (in this case assuming $\frac{m}{\omega_{p}}\ll1$).
\label{fig: conditions on refractive indices}}
\end{figure}

Using the inequalities in Equation~(\ref{eq: condition for the phase matching conditions}) we can explore the range of desirable indices of refraction for a particular setup choice.
 For example, suppose we use monochromatic filter for the signal at half the energy
of the pump photons, $\alpha_\omega=\frac{1}{2}$.
Figure~\ref{fig: conditions on refractive indices} shows in blue the region where the phase matching condition is satisfied for the SPDC process
and in orange the region where the phase matching condition is satisfied
for dSPDC. As can be seen, dSPDC is more restrictive
in the refractive indices. Furthermore, typical SPDC experiments employ
materials such as \emph{beta barium borate }(BBO), \emph{potassium
dideuterium phosphate} (KDP) and \emph{lithium triborate} (LBO)~\cite{kwiat_ultra-bright_1999,couteau_spontaneous_2018,ling_absolute_2008,boeuf_calculating_2000}
that do not allow the phase matching for dSPDC for this example.

A second example which is well motivated is the forward region, namely $\theta_s=0$. In this case phase matching can be achieved so long as 
\begin{equation}
\frac{\left(n_s-n_p\right)^2}{n_s^2-1}\ge \frac{m_\varphi^2}{\omega_p^2}\,,\label{eq:req}
\end{equation}
and $n_s>n_p$.
We will investigate colinear dSPDC in greater detail in~\cite{ourpaper}.

The dSPDC phase matching requirement $n_{s}>n_{p}$ can be achieved in practice by several effects. The most
common one, employed in the majority of SPDC experiments, is birefringence~\cite{moreau_imaging_2019,kwiat_ultra-bright_1999,magnitskiy_spdc-based_2015,boeuf_calculating_2000}.
In this case the polarization of each photon is used to obtain a different
refractive index. For instance, in calcite, the ordinary and extraordinary polarizations have indices of refraction of $n_o=1.658$ and $n_e=1.486$. Taking the former to be the signal and the later for the pump, phase matching can be met for $m_\varphi<0.16 \omega_p$.  Birefringence may also be achieved in single-mode fibers and waveguides with non-circular cross section, e.g. polarization-maintaining optical fibers~\cite{1074847}. Of course, the dependence of the refractive index on frequency may also be used to generate a signal-pump difference in~$n$. 


\subsection{Linear and Birefringent Media for Axions}

Axion electrodynamics is itself a nonlinear theory. Therefore dSPDC with emission of an axion can occur in a perfectly linear medium. The optical medium is needed in order to satisfy the phase matching condition that cannot be satisfied in
vacuum. Because the pump and signal photons in axion-dSPDC have different polarizations, a birefringent material can satisfy the requirement of $n_s>n_p$ and Equation~(\ref{eq:req}).
Because SPDC can be an important background to a dSPDC axion search, using a linear or nearly linear medium, where SPDC is absent is desirable. Interestingly, materials with a crystalline structure that is invariant under a mirror transformation $\VECTOR r\to-\VECTOR r$ will have a vanishing~$\chi^{\PARENTESIS 2}$ from symmetry considerations~\cite{boyd_nonlinear_2008}.

\subsection{Longitudinal Susceptibility for Dark Photon Searches}

As opposed to the axion case, a dSPDC process with a dark photon requires a nonlinear optical medium.
As a result, SPDC can be a background to dSPDC dark photon searches.
In SPDC, the pump, signal, and idler are standard model photons and their polarization is restricted to be orthogonal to their propagation. As a consequence, when one wants to enhance an SPDC process, the nonlinear medium is oriented in such a way that the second order susceptibility tensor $\chi^{(2)}$ can appropriately couple to the transverse polarization vectors of the pump, signal, and idler photons.
The effective coupling between the modes in question is given by
\begin{equation}\label{eq:chi-SPDC}
\chi^{(2)}_\mathrm{SPDC}\equiv \chi^{(2)}_{jkl}\varepsilon^p_j  \varepsilon^s_k \varepsilon^i_l
\end{equation}
with the $\varepsilon$ being the (transverse) polarization vectors for the pump, signal and idler photons. These are spanned by $(1,0,0)$ and $(0,1,0)$ in a frame in which the respective photon is propagating in the $\hat z$ direction.

In dSPDC with a dark photon there is an important difference. Because the dark photon is massive, it can have a longitudinal polarization, $\varepsilon^{A_L'}_l=(0,0,1)$. 
The crystal may be oriented to couple to this longitudinal mode, giving an effective coupling of
\begin{equation}
\label{eq:chi-darkphoton}
\chi^{(2)}_{A_L'}\equiv \chi^{(2)}_{jkl}\varepsilon^p_j  \varepsilon^s_k \varepsilon^{A_L'}_l
\end{equation}
Maximizing this coupling benefits the search both by enhancing signal and reducing background. The background is reduced because the SPDC effective coupling to transverse modes, Equation~(\ref{eq:chi-SPDC}) is suppressed. The signal is enhanced because the coupling to the longitudinal polarization of the dark photon is suppressed by $m_{A'}/\omega$ rather than $(m_{A'}/\omega)^2$~\cite{An:2013yfc, An:2013yua, Graham:2014sha, ourpaper}.

This motivates either non traditional orientation of nonlinear crystals, or identifying nonlinear materials that would usually be ineffective for SPDC. As an example in the first category, any material which is uses for type II phase matching, in which the signal and idler have orthogonal polarization, can be rotated by 90 degrees to achieve a coupling of a longitudinal polarization to a pump and an idler. In the second category, material with $\chi^{(2)}$ tensors that are non-vanishing only in ``maximally non-diaginal'' $\hat x$-$\hat y$-$\hat z$ elements, would obviously be discarded as standard SPDC sources, but can have an enhanced dSPDC coupling.

\section{Dark SPDC Setups and Rates}\label{sec:rates}

A precise estimate  of the rate is beyond the scope of this work, and will be discussed in more detail in~\cite{ourpaper}.
Here we will re-scale known rate formulae in order to examine the dependence of the rate on the geometric factors such as the pump power and beam area, the crystal length, and the area and angle from which signal is collected. 

\begin{figure}
\begin{centering}
\includegraphics[width=8cm]{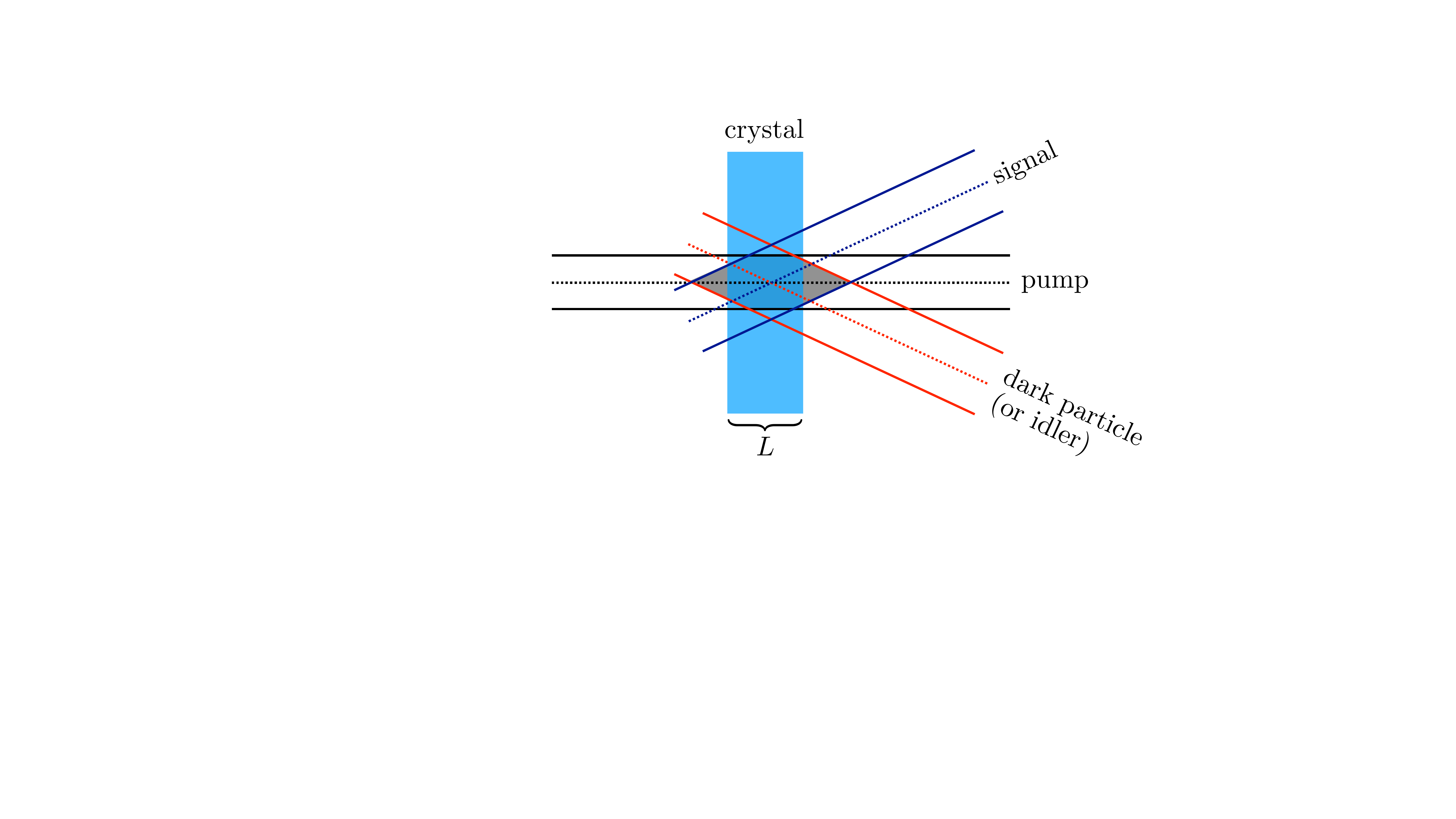}
\par\end{centering}
\caption{A sketch showing the overlap of the pump, signal, and dark particle (or idler) beams with the crystal. 
In general it is desirable to maximize the integration length which is defined by the overlap of the three beams and the crystal.
\label{fig:overlap}}
\end{figure}

Consider the geometry sketched in Figure~\ref{fig:overlap}. A pump beam is incident on an optical element of length~$L$ along the the pump direction. 
The width of the pump beam is set by the laser parameters. The pump beam may consist of multiple modes, or can be guided in a fiber or waveguide~\cite{fiorentino_spontaneous_2007} and in a single mode. 
The width and angle of the signal ``beam'' is set by the apparatus used to collect and detect the signal. This too, can include collection of a single mode in a fiber (e.g.~\cite{ling_absolute_2008,bennink_optimal_2010}) or in multiple modes (as in the CCD example in the next section. 
In SPDC, when the idler photon may be collected the width and angle of the idler can also be set by a similar apparatus. However, in dSPDC (or in SPDC if we choose to only collect the signal) the $\varphi$ beam is not a parameter in the problem. In this case we are interested in an inclusive rate, and would sum over a complete basis of idler beams. Such a sum will be performed in~\cite{ourpaper}, but usually the sum will be dominated by a set of modes that are similar to those collected for the signal.

Generally, the signal rate will depend on all of the choices made above, but we can make some qualitative observations.
The rate for SPDC and dSPDC will be proportional to an integral over the volume defined by the overlap of the three beams in Figure~\ref{fig:overlap} and the crystal. When the length of the volume is set by the crystal, the process is said to occur in the ``thin crystal limit''. In this limit, the beam overlap is roughly a constant over the crystal length and thus the total rate will grow with $L$. Since dSPDC is a rare process, we observe that taking the collinear limit of the process, together with a thicker crystal may allow for a larger integration volume and an enhanced rate.

The total rate for SPDC in a particular angle, integrated over frequencies, in the thin crystal limit is~\cite{ling_absolute_2008, ourpaper},
\begin{equation} \label{eq:rate-tot}
\Gamma_\mathrm{SPDC} \sim 
\frac{  P_p {\chi^{(2)}_\mathrm{eff}}^2 \omega_s \omega_i L}{ \pi  n_p n_s n_i A_\mathrm{eff}} 
\end{equation} 
where $L$ is the crystal length, $P_p$ is the  pump power, $A_\mathrm{eff}$ is the effective beam area, and ${\chi^{(2)}_\mathrm{eff}}$ is the effective coupling of Equation~(\ref{eq:chi-SPDC}). 
A parametrically similar rate formula applies to SPDC in waveguides~\cite{fiorentino_spontaneous_2007,ourpaper}. The pump power here is the effective power, which may be enhanced within a high finesse optical cavity, e.g.~\cite{Ehret:2009sq}.
The inverse dependence with effective area can be understood by since the interaction hamiltonian is proportional to electric fields, which grow for fixed power for a tighter spot. For ``bulk crystal'' setups a pair production rate of order few times $10^6$ per mW of pump power per second is achievable~\cite{bennink_optimal_2010} in the forward direction. In waveguided setups, in which the beams remain confined along a length of the order of a cm, production rates of order $10^9$ pairs per mW per second are achievable in KTP crystals~\cite{fiorentino_spontaneous_2007, spillane_spontaneous_2007}, and rates of order $10^{10}$ were discussed for LN crystals~\cite{spillane_spontaneous_2007}. Also noteworthy are  cavity enhanced setups in which the effective pump power is increased by placing the device in an optical cavity, and/or OPOs, in which the signal photon is also a resonant mode, including compact devices with high conversion efficiencies, e.g~\cite{furst_low-threshold_2010, beckmann_highly_2011,werner_continuous-wave_2015, lu2021ultralowthreshold}.

An important scaling of this rate is the  $L/A_\mathrm{eff}$ dependence. This scaling applies for dSPDC rates discussed below. Within the thin crystal limit one can thus achieve higher rates with: (a) a smaller beam spot, and (b) a thicker crystal. It should be noted that for colinear SPDC, the crystal may be in the ``thin limit'' even for thick crystals (see Figure~\ref{fig:overlap} and imagine zero signal and idler emission angle). 

\subsection{Dark Photon dSPDC Rate}
The dSPDC rate into a dark photon with longitudinal polarization, $A_L'$, can similarly be written as a simple re-scaling of the expression above
\begin{equation} \label{eq:dark-photon-rate-tot}
\Gamma^{(A'_L)}_\mathrm{dSPDC} \sim 
\epsilon^2 \frac{m_{A'}^2}{\omega_{A'}^2}\frac{  P_p {\chi^{(2)}_{A_{L}'}}^2 \omega_s \omega_{A'} L}{  n_p n_s A_\mathrm{eff}} 
\end{equation} 
where the effective coupling $\chi^{(2)}_{A_{L}'}$ is defined in Equation~(\ref{eq:chi-darkphoton}). This is valid
in regions where the dark photon mass is smaller than the pump frequency, such that the produced dark photons are relativistic and have a refractive index of 1. Using the optimistic waveguide numbers above as a placeholder, assuming an optimized setup with similar $\chi^{(2)}$, the number of events expected after integrating over a time $t_\mathrm{int}$ are
\begin{equation}
N^{(A'_L)}_\mathrm{events} \sim 10^{21}\, 
\left(\epsilon^2 \frac{m_{A'}^2}{\omega_{A'}^2}\right) 
\left(\frac{P_p}{\mathrm{Watt}}\right)
\left(\frac{L}{\mathrm{m}}\right)
\left(\frac{t_\mathrm{int}}{\mathrm{year}}\right)\,.
\end{equation}
The current strongest lab-based limit is for dark photon masses of order 0.1~eV is set by the ALPS experiment at $\epsilon\sim 3\times 10^{-7}$.
For this dark photon mass $m_{A'}\sim 0.1\omega_p$,  the $m_{A'}/\omega_{A'}$ term does not represent a large suppression. In this case of order~10 dSPDC events are produced in a day in a 1~cm crystal with a Watt of pump power with the assumptions above. This implies that a relatively small dSPDC experiment with an aggressive control on backgrounds could be used to push the current limits on dark photons.   

Improving the limits from solar cooling, for which $\epsilon m_{A^\prime}/\omega_{A^\prime}$ is constrained to be smaller,  would represent an interesting challenge.  Achieving ten events in a year of running requires a Watt of power in a waveguide greater than 10 meters
(or a shorter waveguide with higher stored power, perhaps using a Fabri-Perot setup). Interestingly, in terms of system size, this is still smaller than the ALPS-II experiment which would reach 100 meters in length and an effecting power of a hundred~kW. This is because  LSW setups require both production and detection, with limits scaling as $\epsilon^4$.

\begin{table*}
\begin{center}
\begin{tabular}{lcc}
\hline\hline
 & \qquad\ Dark Photon ($m_{A'}=0.1$~eV) \qquad\qquad & Axion-like particle ($m_{a}=0.1$~eV) \\ 
 \hline
Current lab limit & $\epsilon < 3\times 10 ^{-7}$  & $g_{a\gamma}<10^{-6}$~GeV$^{-1}$  \\
Example dSPDC setup & $P_p=1$~W & $P_p=1$~kW \\
 & $L=1$~cm   & $L=10$~m \\
 & $\Gamma=10$/day & $\Gamma=10$/day \\
\hline
Current Solar limit & $\epsilon < 10^{-10}$  & $g_{a\gamma}<10^{-10}$~GeV$^{-1}$   \\
Example dSPDC setup & $P_p=1$~W & $P_p=100$~kW \\
 & $L=10$~m & $L=100$~m \\
  & $\Gamma=10$/year & $\Gamma=10$/year \\
\hline\hline
\end{tabular}
\caption{ Current lab-based and Solar-based based limits on the couplings of dark photns and axion-like particles with a benchmark mass of 0.1~eV. For each limit we show the parameters of an example dSPDC in a waveguide and the rate it would produce for couplings that would produce the specified benchmark rate with the corresponding coupling. The pump power is an effective power which can include an enhancement by an optical cavity setup.  For dark photon rates we assume a nonlinearity of the same order found in KTP crystals. }
\end{center}
\end{table*}

\subsection{Axion dSPDC Rate}
A similar rate expression can be obtained for axion-dSPDC 
\begin{equation} \label{eq:axion-rate-tot}
\Gamma^{\mathrm{(axion)}}_\mathrm{dSPDC} \sim 
\frac{  P_p g_{a\gamma\gamma}^2 \omega_s L}{ \omega_{\mathrm{axion}} n_p n_s A_\mathrm{eff}} 
\end{equation} 
where the different scaling with the frequency of the dark field is due to the different structure of the dSPDC interaction (recall that $\chi^{(2)}$ carries a mass dimension of $-2$ while the axion photon coupling's dimension is $-1$). 
Optimal SPDC (dSPDC) rates are acieved in waveguides in which the effective area is of order the squared wavelength of the pump and signal light.
Assuming a (linear) birefringent material that can achieve dSPDC phase matching for an axion the number of signal event scales as
\begin{equation}
N^{(\mathrm{axion)}}_\mathrm{events} \sim 40\, 
\left(
\frac{g_{a\gamma}}{10^{-6}~\mathrm{GeV}^{-1}}\right)^2
\left(\frac{P_p}{\mathrm{Watt}}\right)
\left(\frac{L}{\mathrm{m}}\right)
\left(\frac{t_\mathrm{int}}{\mathrm{year}}\right)\,.
\end{equation}
This rate suggests a dSPDC setup is promising in setting new lab-based limits on ALPs. For example, a 10~meter birefringent single mode fiber with kW of pump power will generate of order 10 events per day for couplings of order $10^{-7}$~GeV$^{-1}$. To probe beyond the CAST limits in $g_{a\gamma}$ may be possible and requires  a larger setup, but not exceeding the scale, say, of ALPS-II. In a 100 meter length and an effective pump power of 100 kW, a few dSPDC signal events are expected in a year.

Maintaining a low background, would of course be crutial. We note, however, that optical fibers are routinely used over much greater distances, maintaining coherence (e.g.~\cite{2020arXiv200711157V}), and an optimal setup should be identified.      

\subsection{Backgrounds to dSPDC}

There are several factors that should be considered for the purpose of reducing backgrounds to SPDC:
\begin{itemize}
    \item Crystal Length and Signal bandwidth:
    In addition to the growth of the signal rate, the signal bandwidth in many setups will decrease with $L$. If this is achieved the signal to background ratio in a narrow band around the dSPDC phase matching solutions will scale as $L^2$. 
    \item Timing:
    The dSPDC signal consists of a single photon whereas SPDC backgrounds will consist of two coincident photons. Backgrounds can be reduced using fast detectors and rejecting coincidence events. 
    \item Optical material: As pointed out in Section~\ref{sec:optimize}, linear birefringent materials can be used to reduce SPDC backgrounds to axion searches. Nonlinear materials with a $\chi^{(2)}$ tensor which couples purely to longitudinal polarizations may be used to enhance dark photon dSPDC events without enhancing SPDC. This technique to reduce SPDC may also be used in axion searches.
    \item Detector noise and optical impurities: Sources of background which may be a limiting factor for dSPDC searches include detector noise, as well as scattering of pump photons off of impurities in the optical elements and surfaces. The Skipper CCD, one example of a detector technology with low noise, will be discussed in the next section. 
\end{itemize}
An optimal dSPDC based search for dark particles will likely consider these factors, and is left for future investigation.
\section{Experimental proof of concept}\label{sec:skipper}

In this section we present an experimental SPDC angular imaging measurement with high resolution employing a Skipper CCD and a BBO nonlinear crystal. In this setup and for the chosen frequencies, dSPDC phase matching is not achievable at any emission angle. Instead, this experiment serves as a proof of concept for the high resolution imaging technique. 

Imaging the dSPDC requires the detection of single photons with low noise and with high spatial and/or energy resolution. A technology that can achieve this
is the Skipper~CCD which is capable of measuring the charge stored in each pixel with single electron resolution~\cite{tiffenberg_single-electron_2017},
ranging from very few electrons (0, 1, 2,~$\dots$) up to more than a thousand (1000, 1001, 1002,~$\dots$)~\cite{Rodrigues2020}. 
This unique feature combined with the high spatial resolution typical of a CCD detector makes this technology very promising for the detection of small optical signals with a very high spatial resolution. 

\subsection{Description of the experimental setup}

\begin{figure}
\begin{centering}
\includegraphics[width=1\columnwidth]{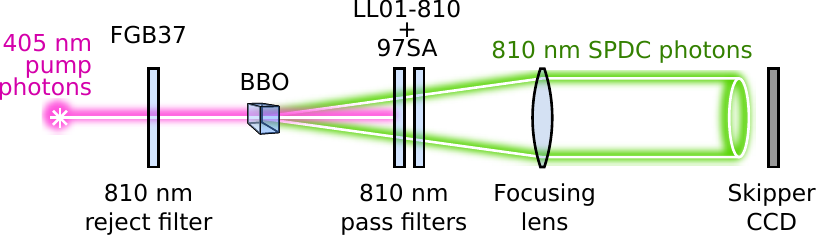}
\par\end{centering}
\caption{Pictorial representation of the experimental setup implemented in
the lab. A $405\protect\NANO m$ diode laser was used as a source
of pump photons. The beam passes through a $810\protect\NANO m$ rejection
filter, then through the BBO crystal, after this through two $810\protect\NANO m$
band pass filters, a focusing lens and finally the SPDC photons reach
the CCD detector. \label{fig:Pictorial-representation-of experiment}}
\end{figure}

\begin{figure}
\begin{centering}
\includegraphics[width=1\columnwidth]{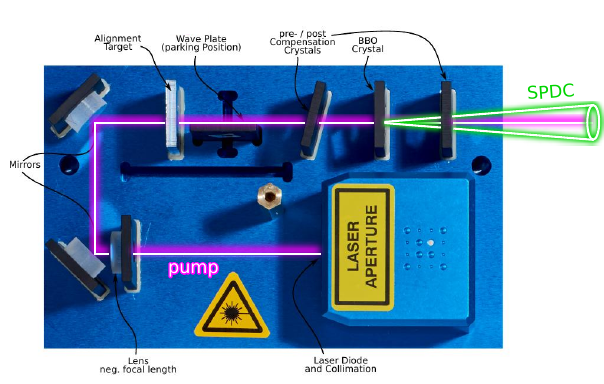}
\par\end{centering}
\caption{Picture of the source of entangled photons which is a part of a commercial
system. The FGB37 filter shown in Figure~\ref{fig:Pictorial-representation-of experiment}
was placed just after the laser aperture. All the other components
were placed outside this device. Figure adapted from~\cite{qutools_qued_nodate}. \label{fig:Picture-of-the quED}}
\end{figure}

With the aim of comparing the developed phase space model against real data,
the system depicted in Figure~\ref{fig:Pictorial-representation-of experiment}
was set up. A source of entangled photons that employs SPDC, which is part
of a commercial system\footnote{\url{https://www.qutools.com/qued/}}, was used. 
Two type I nonlinear BBO crystals are used as a high
efficient source of entangled photons~\cite{kwiat_ultra-bright_1999}.
This pre-assembled source can be seen in Figure~\ref{fig:Picture-of-the quED}.
However, in this work, we did not take advantage of the polarization
entanglement, we only use the energy-momentum conservation to get
spatially correlated twin photons. In addition to the original design,
we added a Thorlabs FGB37 filter after the laser to remove a small
$810\NANO m$ component coming along with the pump beam. After the
BBO crystal, where the SPDC process occurs, two additional $810\NANO m$
band pass filters (one Semrock Brighline LL01-810 and one Asahi Spectra
97SA) were placed to prevent the $405\NANO m$ pump beam from reaching
the detector. We also placed a lens at its focal distance from the
BBO to prevent the SPDC cone from spreading during its travel to the
CCD (see Figure~\ref{fig:Pictorial-representation-of experiment}).
Thus, we got the SPDC ring projected on the CCD surface with a radius
of $110\UNIT{pixel}$, which corresponds to $1650\MICRO m$. This
setup is also part of a research where the novel features of Skipper CCD~\cite{tiffenberg_single-electron_2017}
are being tested for Quantum Imaging. The capability of Skipper-CCD
to reduce the readout noise as low as desired taking several samples
of the charge in each pixel was not used in this work to acquire data, but it was for the calibration. Details about
the same detector used here can be found in reference~\cite{Rodrigues2020}.

We used an absolute calibration between the Analog to Digital Units (ADU)
measured by the amplifier from the CCD and the number of electrons
in each pixel, possible with the Skipper-CCD,
which was previously carried out for this system~\cite{Rodrigues2020}.
Thus, we reconstructed the number of photons per pixel, since at $810\NANO m$
it can be assumed that one photon creates at most one electron, by reason
of its energy is $0.43\UNIT{eV}$ above the silicon band gap
($1.1\UNIT{eV}$). The factor affecting that relationship is
the efficiency which -besides being very high ($\sim 90\UNIT{\%}$) at
this wavelength- is pretty uniform over the entire CCD surface~\cite{Bebek_2017}. 

\subsection{Results}
Using the setup previously described, 400 images with $200\UNIT s$ exposure each 
were averaged to produce the data used for comparison with the model. 
Thus, we reduced a factor twenty the uncertainty
in the expected number of photons in each pixel and drastically reduced
the dark counts coming from random background. This significantly
improved the identification of minima between rings, which results
to be crucial to compare the experimental data with the presented theory.

\begin{figure}
\begin{centering}
\includegraphics{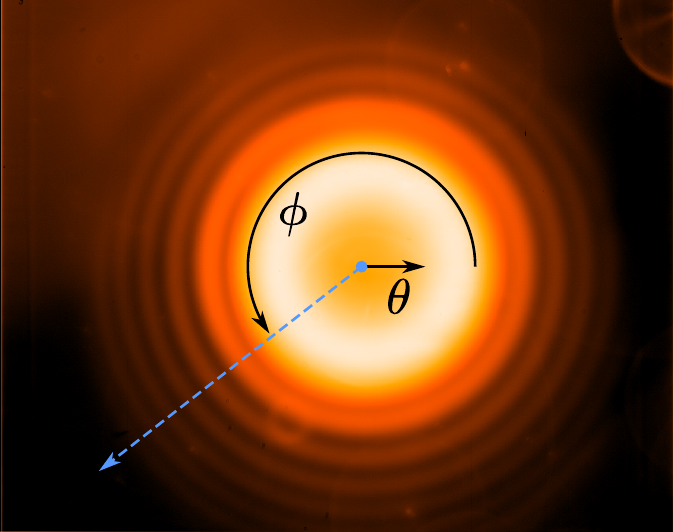}
\par\end{centering}
\caption{Average of 400 images with $200\protect\UNIT s$ exposure each to the SPDC
photons each using the setup depicted in Figure~\ref{fig:Pictorial-representation-of experiment}.
The angular coordinates $\theta$ and $\phi$ are indicated in the
figure. The blue dashed line was used to extract the data profile shown
in Figure~\ref{fig: SPDC profile plot data vs model}. \label{fig:Average-of-400 images FOTO DEL ANILLO PAPA}}
\end{figure}

Figure~\ref{fig:Average-of-400 images FOTO DEL ANILLO PAPA} presents
the averaged image of the SPDC ring coming out of the BBO crystal.
The main ring is clearly visible, as well as many secondary
maxima and minima. The angular coordinates $\theta$ and $\phi$ were
indicated on the image. It has to be noted that the angular coordinates
used in the previous sections are the ones inside the optical medium.
Since the CCD detector is outside the BBO crystal, Snell's law must
be used to relate the angles inside and outside the nonlinear crystal. 
Specifically, for the $\theta$ coordinate this is 
\[
\theta=\arcsin\PARENTESIS{\frac{\sin\Theta}{n}}
\]
 where $\theta$ is the angle inside the optical medium in which
the SPDC process occurs, $\Theta$ is the angle outside the optical
medium and $n$ is the refractive index. In this work we apply the transformation to refer all angles to
those inside the optical medium in which the interaction occurs. 

\begin{figure}
\begin{centering}
\includegraphics[width=1\columnwidth]{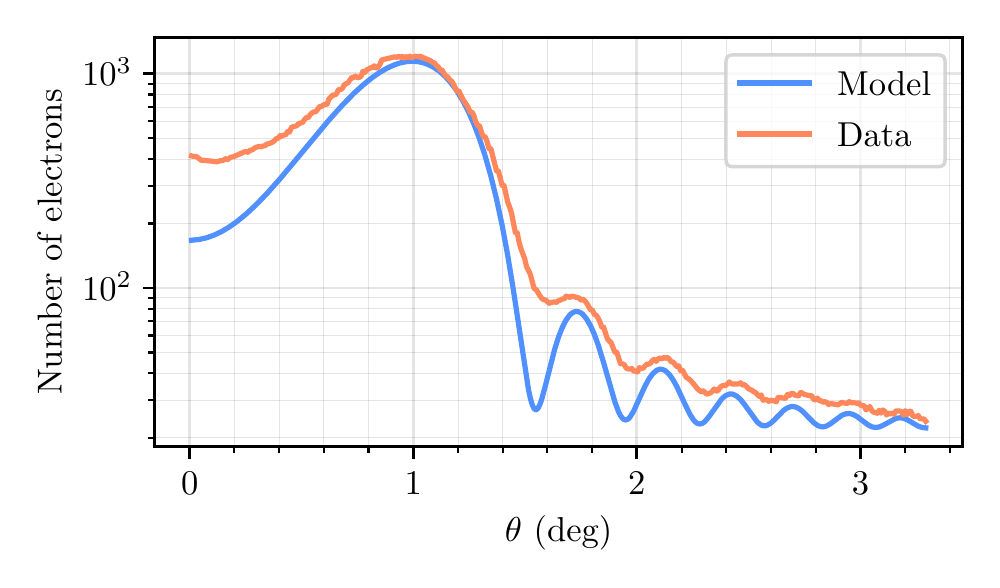}
\par\end{centering}
\caption{Experimental intensity profile for the SPDC ring along the blue dashed
line in Figure~\ref{fig:Average-of-400 images FOTO DEL ANILLO PAPA}
compared with theoretical distribution. The $y$ axis scale is the number of electrons in each pixel of the CCD detector obtained via the absolute Skipper-CCD calibration described. \label{fig: SPDC profile plot data vs model}}
\end{figure}

As can be seen in the image, even though it is possible to see many
of the maxima and minima, there is a non uniform background component
produced by reflections and other imperfections of the experimental configuration. The current implementation
of our setup has some mechanical limitations that make hard to isolate, measure
and remove this non-uniform background component. Still, this image
can be used to compare the theory with the experiment, so the profile
of the intensity distribution along the blue dashed line shown in
Figure~\ref{fig:Average-of-400 images FOTO DEL ANILLO PAPA} was
extracted and plotted against the model. 
This plot is shown in Figure~\ref{fig: SPDC profile plot data vs model}. Some comments about this plot:

\begin{itemize}
    \item A uniform background component was added to the model.
    \item The transmittance curve of the $810 \NANO m$ filters provided by the manufacturers was taken into account.
\end{itemize}

So, to be specific, the plot
in Figure~\ref{fig: SPDC profile plot data vs model} uses as model
the intensity profile given by
\[
I_{\text{model}}\PARENTESIS{\theta} = I_0 \int\frac{d^2 \hat{\Gamma}}{d(\cos\theta_s) d\alpha_\omega} F(\alpha_\omega) d\alpha_\omega + I_1
\]
 where $I_0$ and $I_1$ are real numbers fitted, $\frac{d^2 \hat{\Gamma}}{d(\cos\theta_s) d\alpha_\omega}$ is given by Equation~(\ref{eq: F_alpha_theta_phi}) and $F(\alpha_\omega)$ is the transmittance of the filters as provided by the manufacturers, which is a sharp peaked function centered at $810 \NANO m$ with a pass band of $10 \NANO m$ width.

For the parameters of the distribution $\hat{\Gamma}$, we used:
\begin{itemize}
\item $405\NANO m$ for the pump wavelength as provided by the manufacturer
of the laser, 
\item the refractive indices for the BBO crystal were taken from \url{https://refractiveindex.info/}~\footnote{The extraordinary refractive index was taken from
\url{https://refractiveindex.info/?shelf=main&book=BaB2O4&page=Tamosauskas-e}
(Tamo¨auskas et al., 2018) and the ordinary refractive index from
\url{https://refractiveindex.info/?shelf=main&book=BaB2O4&page=Eimerl-o}
(Eimerl et al., 1987).} 
\item and the crystal length was assumed to be $L=1.14\MILI m$ after
fine tuning~\footnote{It was not possible to obtain precise information of the dimensions
of the crystal from the manufacturer. Furthermore, the length of the
crystal is not easy to measure with our current implementation without
breaking the SPDC source shown in Figure~\ref{fig:Picture-of-the quED}.
So we decided to use a ``reasonable value'' and fine tune it.}. 
\end{itemize}
We had to perform a fine tuning of $n_s - n_p$ too (since signal and idler have the same frequency and angle then $n_s=n_i$). The model is very sensitive to this difference,
and this quantity depends on factors such as the temperature of the
BBO and its precise orientation in space, which were not measured.
So it is not surprising that we had to perform this fine tuning on~$n_s-n_p$. The final values used for the refractive indices were $n_p \approx 1.66082$ and $n_s = n_i \approx 1.66107$.
Although this fine tuning had to be done, all parameters are very well within expected values.

As seen both in Figures~\ref{fig:Average-of-400 images FOTO DEL ANILLO PAPA}
and~\ref{fig: SPDC profile plot data vs model} the phase space
factor, studied in previous sections, is the dominant modulation in the distribution
of photons for the SPDC process in our setup. Even though there is a non negligible
background component, it is evident the dependence $I\sim\SINC^{2}\PARENTESIS{\theta^{2}}$
for small values of~$\theta$ predicted by Equation~(\ref{eq: F_alpha_theta_phi}).

\section{Discussion}\label{sec:conclusions}

We have presented a new method to search for new light and feebly coupled particles, such as axion-like particles and dark photons. The dSPDC process allows to tag the production of a dark state as a pump photon down-converts to a signal photon and a ``dark idler'' $\varphi$, in close analogy to SPDC. We have shown that the presence of 
indices of refraction that are different than~1 open the phase space for the decay, or down conversion, of the massless photon to the signal plus the dark particle, even if $\varphi$ has a mass. This type of search has a parametric advantage over light shining through wall setups since it only requires producing the axion or dark photon, without a need to detect it again. Precise sensitivity calculations for dSPDC with Dark Photon and Axion cases that can be achieved through this method are ongoing~\cite{ourpaper}.

The commonplace use of optics in telecommunications, imaging and in quantum information science, as well as the development of advanced detectors, can thus be harnessed to search for dark sector particles. Increasing dSPDC signal rates will require high laser power,  long optical elements. 
Enhancing the signal and suppressing SPDC backgrounds also requires identifying the right optical media for the search. Axion searches would benefit from optically linear and birefringent materials, with greater birefringence allowing to search for higher axion masses. Searches for dark photons would benefit from strongly nonlinear materials that are capable of coupling to a longitudinal polarization. This, in turn, motivated either non-conventional optical media, or using conventional crystals that are oriented by a $90^\circ$ rotation from that which is desired for SPDC. 
Enhancing the effective pump power with an optical cavity is a straightforward way to enhance the dSPDC rate. We leave the exploration of a ``doubly resonant'' OPO setup in which the signal is also a cavity eigenmode (in parallel with~\cite{Ehret:2009sq}) for future work.  
Finally, detection of rare signal events requires sensitive single photon detectors with high spatial and/or frequency resolution.

We also performed an experimental demonstration of the one of the setups discussed using a Skipper CCD for SPDC imaging. The setup we used for this demonstration was adapted from one designed to enforce SPDC and thus does not open dSPDC phase space. We show that the Skipper technology 
allows one to measure with high accuracy. Thus, quantum teleportation methods or ghost imaging of dark sector particles achieved through phase space engineering is a very promising technique for future explorations of dark sector parameter spaces.
\begin{center}
    \small{$\mathghost\mathghost\mathghost\mathghost$}
\end{center}
\noindent {\bf Acknowledgments:}   We would like to thank  Joe Chapman, Paul Kwiat, and Neal Sinclair for informative discussions. This work was funded by a DOE QuantISED grant.

\bibliographystyle{unsrt}
\bibliography{zotero, non_automatic, citations-3}
 
\end{document}